\documentclass[aps,amsmath,twocolumn,amssymb,floatfix,showpacs,superscriptaddress,footinbib,longbibliography,prl,reprint]{revtex4-2}
\pdfoutput=1
\usepackage{graphicx}
\usepackage{bm}
\usepackage{amsmath}
\usepackage{xcolor} 
\usepackage{braket}
\usepackage{hyperref}
\usepackage{float}
\usepackage[caption=false]{subfig} 

\hypersetup{
    colorlinks=true,
    linktoc=page,
    linkcolor=red,
    citecolor=blue,
    urlcolor=magenta
}
\newcommand{\Nsat}{N_{\rm sat}}

\newcommand{\Urot}{U_{\mathrm{rot}}}
\newcommand{\Uint}{U_{\mathrm{int}}}

\begin{document}

\title{Quantum state revivals, time quasicrystals, and entanglement steering in a Fibonacci-driven central-spin model}

\author{Anushka Agrawal}
\altaffiliation{These authors contributed equally to this work.}
\affiliation{Department of Physics, BITS Pilani, Hyderabad Campus, Telangana 500078, India}
\author{Adithya A. Vasista}
\altaffiliation{These authors contributed equally to this work.}
\affiliation{Department of Physics, BITS Pilani, Hyderabad Campus, Telangana 500078, India}
\author{Tanay Nag}
\affiliation{Department of Physics, BITS Pilani, Hyderabad Campus, Telangana 500078, India}
\author{Sayan Choudhury}
\email{sayanchoudhury@hri.res.in}
\affiliation{Harish-Chandra Research Institute, Chhatnag Road, Jhunsi, Prayagraj (Allahabad) 211019, India}
\affiliation{Homi Bhabha National Institute, Training School Complex, Anushakti Nagar, Mumbai 400094, India}

\date{\today}

\begin{abstract}
Motivated by the search for non-equilibrium phases of matter beyond the Floquet paradigm, we propose a route to realize perfect revivals and time quasicrystals (TQCs) in a Fibonacci-driven central-spin model with Ising interactions between the central and satellite spins in the absence of disorder. We analytically demonstrate that when the interaction strength, $\lambda$ is tuned to $2 m \pi$, no entanglement is generated between the central and satellite spins at stroboscopic times, independent of the strength of the transverse driving field and the drive sequence. This leads to indefinitely long-lived quasiperiodic oscillations in satellite spin magnetization. Furthermore, for initial states with a definite bath magnetization, no entanglement is generated when the transverse driving field, $\theta$ is tuned to $\pi$ for any value of $ \lambda$, and the satellite spin magnetization oscillations exhibit period-3 revivals at Fibonacci times. Examining the robustness of these oscillations away from this fine-tuned limit, we identify a parameter regime where the quasiperiodic response does not decay over the timescales probed, giving rise to a long-lived $\pi$-pulse-proximate TQC regime. We further show that at special Clifford points, the system exhibits perfect quasiperiodic oscillations of the satellite spin magnetization, with a parity-dependent Fourier response. This response persists over the timescales probed, even away from the Clifford points for an even number of satellite spins, leading to the formation of a Clifford-proximate TQC. Finally, we demonstrate that at this Clifford points, the system is steered along an entangled trajectory, leading to the dynamical generation of Bell-cat and spin-cat states. These states exhibit magnetization of collective spin fluctuates quadratically with the number of satellite spins, thereby enabling Heisenberg-limited metrology.
\end{abstract}

\maketitle
\def\thefootnote{*}\footnotetext{These authors contributed equally to this work}

\emph{Introduction:} The dynamics of driven quantum systems have attracted significant attention in recent years~\cite{bukov2015universal,goldman2014periodically,oka2019floquet,sen2021analytic,Weitenberg2021,zhan2024perspective,rudner2020band,sato2025floquet}. These systems provide a pathway to realize intrinsically non-equilibrium phases of matter such as discrete time crystals~\cite{sacha2018time,else2020discrete,khemani2019brief,zaletel2023colloquium} and Floquet topological phases~\cite{rudner2013anomalous,nathan2019anomalous,qiu2026observation,li2023floquet,nag2021anomalous, PhysRevB.105.155406,ghosh2024generation,prnx-47mk}. Furthermore, driving protocols have important applications in the development of quantum technologies, including state preparation and dynamical control~\cite{PhysRevLett.123.090602,3dpp-p2pr,Gangopadhay_2025,PhysRevB.104.174308,PhysRevB.100.134301,PhysRevB.89.165425}, sensing \cite{PhysRevA.109.L050203, PhysRevB.111.024315, sharma2026quantummetrologypartiallyaccessible,fiderer2018quantum}, communication \cite{PhysRevResearch.6.013116}, and simulation~\cite{geier2021floquet,katz2025floquet}. 

While most studies on driven systems initially focused on the case of periodic driving, recent work has extended this focus to the domain of quasi-periodic driving~\cite{verdeny2016quasi,else2020long,gallone2024prethermalization,anisur2026dissipativedicketimequasicrystals,anisur2025quasi,PhysRevLett.120.070602,dumitrescu2022dynamical,schmid2025self,dutta2025prethermalization,prg5-3srn,PhysRevB.104.134307,lhnv-1l4l}. These studies have unveiled an array of intriguing phenomena -- ranging from topological energy conversion~\cite{martin2017topological,peng2018topological,qi2021universal,prg5-3srn} to non-equilibrium phases such as time quasicrystals. However, due to the presence of multiple incommensurate frequencies in the drive, these systems generically heat up to an infinite temperature state. Thus, stabilizing these systems and controlling the generation of multipartite entanglement is inherently challenging which we address in this work.

In this Letter, we demonstrate that collective spin models provide a powerful platform to realize time quasicrystalline phases in the absence of disorder. To this end, we consider the paradigmatic central-spin model (CSM)~\cite{PhysRevLett.96.140604,PhysRevLett.120.070602,t7m4-1qc4,PhysRevB.86.020401,PhysRevA.93.012112,PhysRevE.93.062119,PhysRevA.72.052113}, composed of a central spin-$1/2$ object interacting with $N_{\rm sat}$ satellite spins via Ising interactions, under a quasiperiodic Fibonacci modulation of an external transverse field. First, we identify two parameter lines, where no entanglement is generated between the central and satellite spins at stroboscopic times, leading to indefinitely long-lived quasiperiodic oscillations of the magnetization. Notably, these outcomes remain unchanged for other binary quasiperiodic driving sequences. These oscillations persist up to long times, for a finite parameter window around one of these parameter lines, leading to the formation of a prethermal $\pi-$pulse-proximate time quasicrystal (TQC). Furthermore, we demonstrate that at the Clifford points, the system exhibits indefinitely long-lived quasi-periodic oscillations of both the magnetization and the entanglement entropy. The Fourier response of the system at these points is dictated by the parity of the satellite spins, with a Clifford-proximate TQC emerging for even $N_{\rm sat}$. Finally, we demonstrate that the entangled states generated at the Clifford points exhibit a quadratic scaling of the variance associated with the collective spin magnetization, thereby providing a pathway towards Heisenberg-limited metrology.

\begin{figure}
    \centering
    \includegraphics[width=0.47\textwidth]{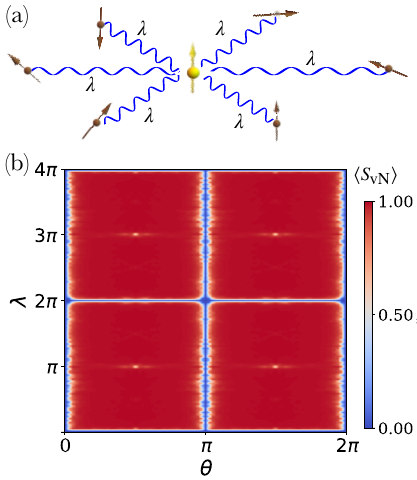}
    \caption{{\bf Model and Phase diagram:} (a) Schematic illustration of the central spin model. (b) Time-averaged entanglement entropy ($S_{\rm vN}$), evaluated up to $2000$ drive periods for $\Nsat=200$. The system does not exhibit any entanglement growth when $\theta = n \pi \, \forall \lambda$ and $\lambda = 2 n \pi \, \forall \theta$. Furthermore, non-ergodic behavior with an intermediate value of $S_{\rm vN} \, (0< S_{\rm vN}<1)$ is observed for Clifford points $(\lambda = (2n+1) \pi, \, \theta = (2m+1)\pi/2)$. Most of the remaining parameter space is characterized by $S_{\rm vN} \sim 1$, consistent with chaotic dynamics. Here $n,m \in \mathbb{Z}$}
    \label{fig:time-avg entropy}
\end{figure}

\emph{Model and dynamical regimes:} We examine the dynamics of CSM, subjected to transverse field kicks of strength $\theta$ at $t_m = \big\lfloor \phi m \big\rfloor$ with  $\phi= (\sqrt{5}+1)/2$:
\begin{equation}
H(t) = -\lambda S_c^z \sum_{i=1}^{N_{\text{sat}}} S_i^z  +  \theta \sum_{m=1}^{\infty} \Big[ {\sum_{i=1}^{N_{\text{sat}}}S_i^x }+ S_c^x\Big] \delta(t - t_m),
\end{equation}
where, $S^\gamma = \frac{1}{2}\sigma^\gamma$ and $\sigma^\gamma$ are the Pauli matrices. A schematic illustration of
the model is shown in Fig~\ref{fig:time-avg entropy}(a). This model constitutes a generalization of the Floquet protocol studied in Ref.~\cite{biswas2025floquetcentralspinmodel}, and it can be realized in quantum dots~\cite{RevModPhys.85.79}, ultracold atomic gases~\cite{PhysRevA.100.043618,PhysRevA.108.052618}, nitrogen-vacancy centers in diamond~\cite{Schwartz_2018,childress2006coherent}, and nuclear magnetic resonance quantum processors~\cite{PhysRevLett.127.080401}. 

\begin{figure} 
    \centering
   \includegraphics[width=0.5\textwidth]{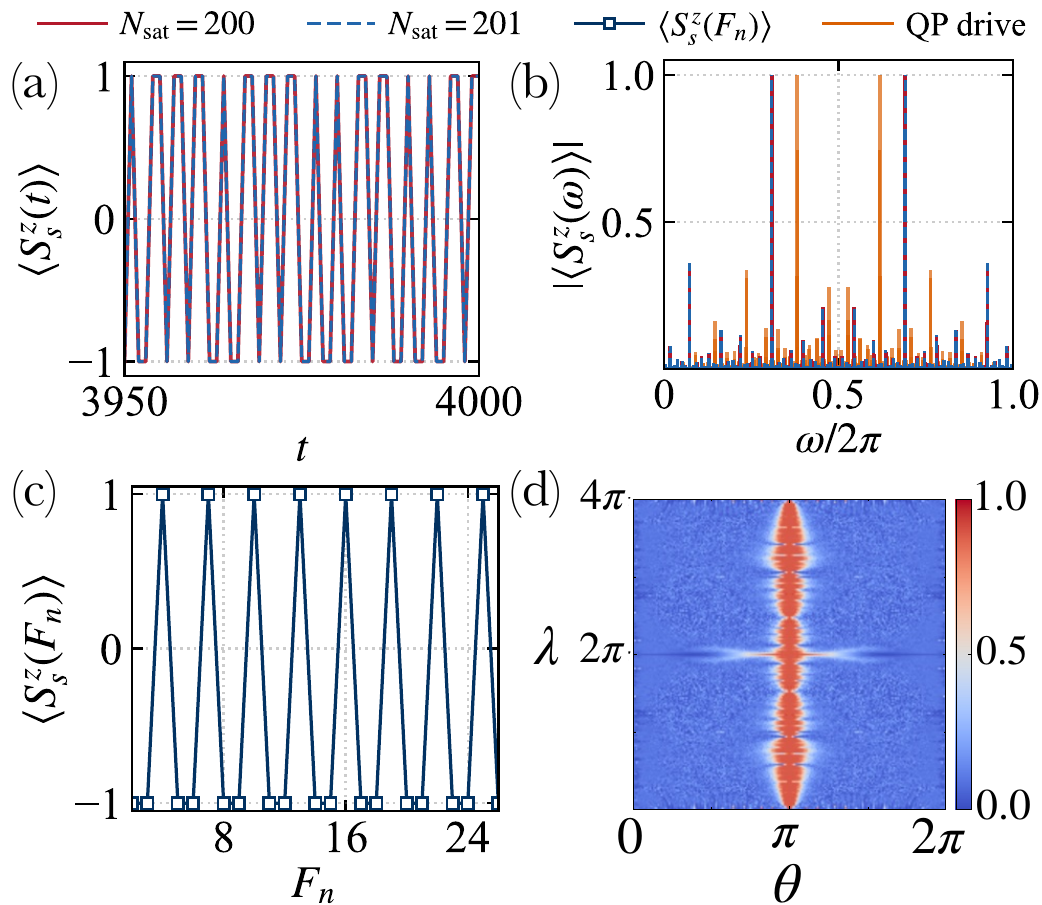}
    \caption{\textbf{Time quasicrystal:} (a) The time evolution of the satellite spin magnetization, $\langle S^z_s \rangle $ under a perfect $\pi$-pulse ($\theta=\pi$). (b) The Fourier spectrum of the magnetization oscillations reveal that the response is shifted sharply from that of the drive. (c) The quasiperiodic magnetization oscillations at $\theta = \pi$ corresponds to period-3 oscillations in Fibonacci time. (d) The time quasi-crystal (TQC) fraction, $f_{\rm TQC}$ [Eq.~(\ref{eq:TQCfrac})], obtained from the Fourier response up to 4000 drive periods for $\Nsat=200$. The system hosts a finite parameter regime where $f_{\rm TQC} \approx 1$ over a finite parameter window, indicating the existence of a long-lived $\pi-$pulse-proximate TQC regime.}
    \label{Fig:pi-pulse}
\end{figure}

In this work, we focus on the stroboscopic dynamics of these systems. Consequently, in the following analysis, we examine the dynamics of the system under a quasiperiodic sequence of two unitaries,
\begin{equation}
    \Uint = \exp\bigg(i\lambda S_c^z \sum_{i=1}^{N_{\text{sat}}} S_i^z\bigg),\,\, \Urot =  \exp\bigg(-i \theta \Big[ {\sum_{i=1}^{N_{\text{sat}}}S_i^x }+ S_c^x\Big]\bigg)
\end{equation}
Furthermore, due to the quasiperiodic nature of the drive, we can obtain the properties of the system at the $n-$th Fibonnaci time, $t=F_n$, where $F_n= (\phi^n - (-1)^n \phi^{-n})/\sqrt{5}$, by using the Unitary, $U_n = U_{n-2}U_{n-1}$, where $U_{0} = U_{\rm rot}$ and $U_1=U_{\rm int}$ and $n \ge 2$. As we discuss below, the TQC phases are characterized by a salient Fibonacci time oscillation period. We analytically obtain these oscillation periods and characterize their parity dependence, thereby taking a significant step beyond the Floquet protocols studied earlier in this setup~\cite{biswas2025floquetcentralspinmodel,PhysRevB.111.014302}

To characterize the distinct dynamical regimes that emerge in this system, we first analyze the dynamics of the entanglement generated between the central and satellite spins. To this end, we construct the reduced density matrix of the central qubit by tracing out the satellite bath, $\rho_c(t) = \text{Tr}_{\text{sat}}[|\psi(t)\rangle\langle\psi(t)|]$, and compute the entanglement entropy: $S_{\rm vN}(t) = -\text{Tr}[\rho_c(t)\log_2\rho_c(t)]$ at stroboscopic times. Our results for the time-averaged entanglement entropy are shown in Fig. \ref{fig:time-avg entropy}(b). Much of the parameter space is characterized by high entanglement generation, which is associated with local chaos in collective spin models~\cite{PhysRevE.99.032213}. Notably, however, there is no entanglement generated when $\theta \approx 0, 2 \pi  \, \forall \lambda$ (dynamical freezing), $\theta \approx \pi \, \forall \lambda$ (time quasicrystal regime), and $\lambda = 2 \pi \, \forall \theta$ (quasiperiodic oscillatory regime). Furthermore, we find that the generation of entanglement is suppressed to an intermediate value when $\lambda = (2n+1) \pi, \theta = (2m+1) \pi/2 \, \forall m,n \in \mathbb{Z}$. These points are associated with Clifford-induced oscillations, leading to the generation of metrologically useful Bell-cat and spin-cat states. We now proceed to analyze these regimes in greater detail.
{\emph {Time quasicrystal regime:}} We start our analysis by examining the time evolution of the system when $\theta = \pi$, corresponding to a perfect $\pi$-pulse. In this case, the satellite spin magnetization, $\langle S^z_s (t) \rangle= 2 \bigg(\sum_{i=1}^{N_{\rm sat}}\langle\psi(t)| S^z_i |\psi(t) \rangle\bigg)/N_{\rm sat}$, exhibits infinitely long-lived quasiperiodic oscillations [see Fig.~\ref{Fig:pi-pulse} (a)], leading to a quasiperiodic response that is sharply distinct from that of the drive [Fig.~\ref{Fig:pi-pulse} (b)]. Crucially, despite the quasiperiodic nature of these oscillations, we observe that the system exhibits perfect period-3 oscillations at Fibonacci times [Fig.~\ref{Fig:pi-pulse} (c)]. These oscillations originate from the fact that at the Fibonacci times, $t=F_n$ the spins have been flipped $F_{n-1} \, {\rm mod} \, 2$ times from their initial state. Consequently, the magnetization oscillations at Fibonacci times occur with the Pisano period, $\pi(2) =3$~\cite{PhysRevLett.120.070602}. We note, however, that these perfect revivals alone are not sufficient to guarantee a prethermal TQC regime characterized by a rigid long-lived quasiperiodic response, even away from $\theta=\pi$.

To characterize whether this system indeed hosts a TQC regime, we compute the TQC fraction~\cite{PhysRevX.15.011055}:
\begin{equation}
    f_{\rm TQC} = \frac{|S(\nu^{*})|}{\sum_{\nu^{*}-\delta}^{\nu^{*}+\delta} |S(\nu)|}
\label{eq:TQCfrac}
\end{equation}
where $\nu^{*}$ denotes the frequency of the dominant Fourier peak corresponding to the TQC order and $\delta$ is set to $0.01$. In this case $\nu^{*}$ is obtained from the Fourier response at $\theta = \pi$.  We find that $f_{\rm TQC}\sim 1$ for satellite spin magnetization in a finite parameter regime around $\theta = \pi$ over the timescales probed, thereby indicating the existence of a robust TQC regime, see Fig.~\ref{Fig:pi-pulse} (d). We note, however, that away from $\theta = \pi$, these TQCs are expected to be prethermal in nature~\cite{PhysRevLett.120.070602,Mori2022heating,zhou2026emergentselfsimilarquantumrevivals}; this heating may be mitigated by employing dissipation~\cite{anisur2026dissipativedicketimequasicrystals}.

\emph{Non-entangling oscillatory regime:} We now analyze the $\lambda = 2 m \pi$ line where no entanglement is generated for all values of $\theta$ and $m \in \mathbb{Z}$.  We begin by noting that for  $\lambda=2m\pi$, evolution under $U_{\rm int}$ over one interval reduces to a product of single-spin operators,
\begin{eqnarray}
  U_{\rm int}\big|_{\lambda=2\pi}
  &=& \exp\!\Big[\tfrac{i\pi}{2}\,\sigma^{z}_{c}\textstyle\sum_{i}\sigma^{z}_{i}\Big]
  = \prod_{i=1}^{\Nsat}\big(i\,\sigma^{z}_{c}\sigma^{z}_{i}\big)\nonumber \\
  &=& i^{\Nsat}\,(\sigma^{z}_{c})^{\Nsat}\prod_{i=1}^{\Nsat}\sigma^{z}_{i}.
  \label{eq:factorize}
\end{eqnarray}
Consequently, no entanglement is generated by $U_{\rm int}$. Furthermore, the transverse kick $U_{\rm rot}$ is likewise a product of single-spin rotations for any $\theta$. Every unitary in the drive is therefore local, and no entanglement is generated, independent of $\theta$. This accounts for the deep blue bands of Fig.~\ref{fig:time-avg entropy} (b).We note that although the satellite spin magnetization exhibits indefinitely long-lived quasiperiodic oscillations throughout this regime, their frequency remains locked only within a window around $\theta=\pi$ and shifts once $\theta$ is detuned beyond it.
Consequently, the TQC does not persist along the entire line
[see Fig.~\ref{Fig:pi-pulse}(d)]. Finally, we note that the argument presented here only employs the form of the $U_{\rm int}$ and $U_{\rm rot}$ and not the order in which they are applied. Thus, no entanglement is generated when $\lambda = 2m \pi\, \forall \theta$ for any binary periodic or aperiodic drive sequence built from these two unitaries. A more detailed analysis of the dynamics is presented in the Supplemental Material~\cite{suppmat}.

\begin{figure*}
    \centering
    \includegraphics[width=\textwidth]{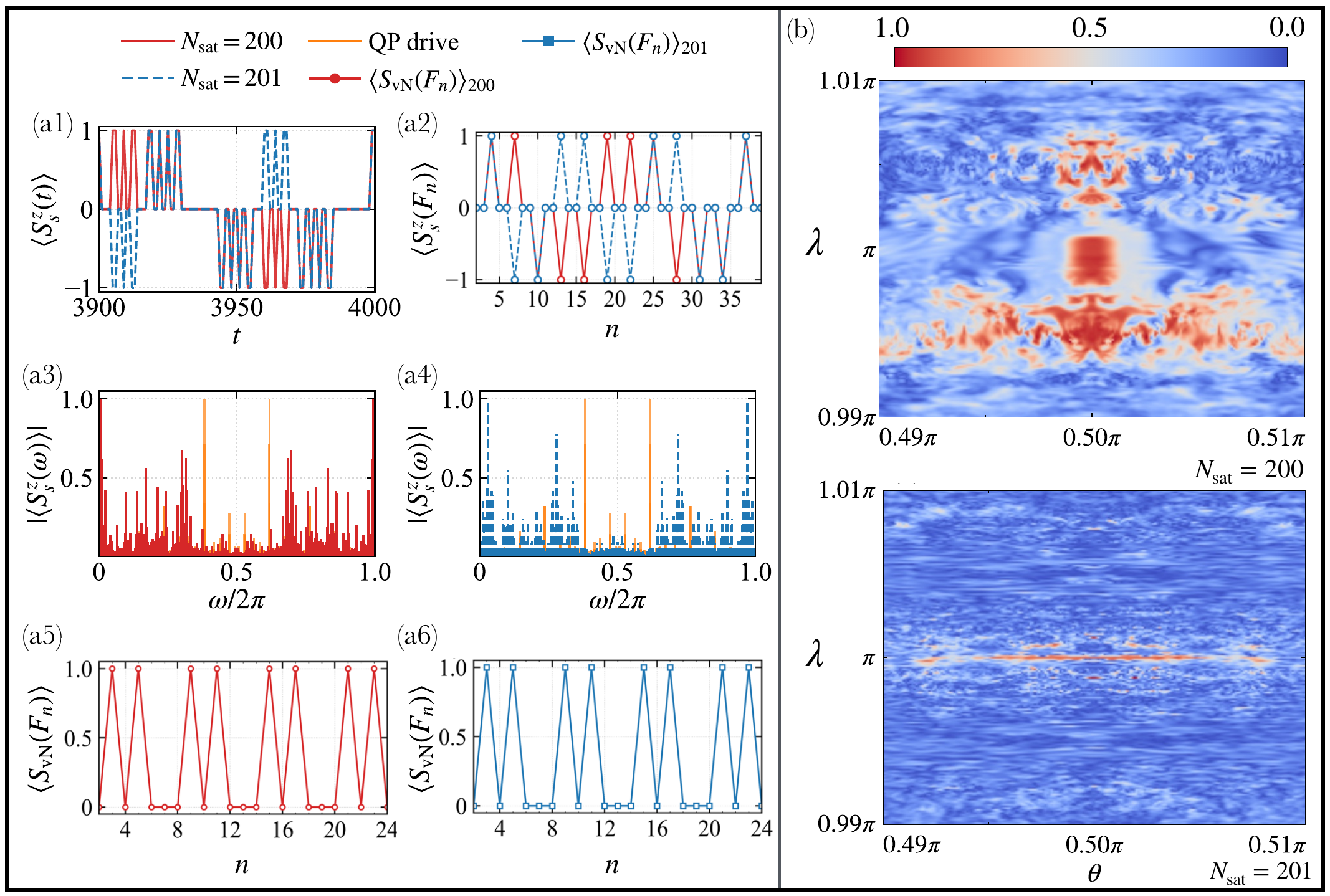}
    \caption{\textbf{Clifford-point dynamics ($\lambda = \pi, \, \theta  = \pi/2$):} (a1)-(a2) The satellite-spin magnetization at long times $\langle S_s^z(t) \rangle$ for even ($N_{\rm{sat}}=200$, solid red) and odd ($N_{\text{sat}}=201$, dashed blue) in the interval $t \in [3900, 4000]$ and at Fibonacci times respectively. The system exhibits quasiperiodic oscillations with a pattern that depends only on the parity of $N_{\rm sat}$. This leads to oscillations in Fibonacci times with a characteristic period of 12 and 18 respectively for odd and even $N_{\rm sat}$. (a3)-(a4) The Fourier spectrum $|\langle S_s^z(\omega) \rangle|$ highlighting the distinct sub-harmonic resonant frequencies distinguishing the two parities. (a5)-(a6) The entanglement entropy, $S_{vN}$ takes only two values during the dynamics (0 and 1) and exhibits period-6 oscillations in Fibonacci times for both even and odd $N_{\rm sat}$. (b) The stability of the Clifford induced quasiperiodic oscillations away from the Clifford points characterized by $f_{\rm TQC}$. The top and bottom panels correspond to $N_{\rm sat} = 200$ and $201$ respectively. The oscillations exhibit greater rigidity for even $N_{\rm sat}$, indicating the existence of a Clifford-proximate TQC regime for even $N_{\rm sat}$.}
    \label{fig:clifford}
\end{figure*}

\emph{Dynamics at the Clifford Points:} We now proceed to examine the time evolution of the system at the Clifford points. We focus our analysis on $\theta=\pi/2$, $\lambda=\pi$, though these results can be generalized to any Clifford point.  At these parameter points, the unitaries $U_{\rm int}$ and $U_{\rm rot}$ map Pauli operators to Pauli operators, and it is possible to analytically track the evolution of any initial stabilizer state, such as the $z$-polarized state considered here.

Due to the permutation symmetry of the satellite spins, the stabilizer generator for the central spin takes the form
$\mathcal{G}_{c}=(-1)^{a}\sigma^{\alpha}_{c}\prod_{i}\sigma^{\beta}_{i}$, and those of the the satellite spins take the form 
$\mathcal{G}^{(j)}_{s}=(-1)^{b}\sigma^{\gamma}_{c}\sigma^{\delta}_{j}
\prod_{i\neq j}\sigma^{\epsilon}_{i}$ throughout the evolution, with
all Pauli indices $\alpha,\beta,\delta,\gamma,\epsilon \in \{0,x,y,z\}$. The generators therefore take values in a finite alphabet whose size is independent of $\Nsat$, on which each drive unitary acts as a fixed permutation. Permutation
symmetry collapses every satellite string to a single power, so
$\Nsat$ can enter only through $(-1)^{\Nsat}$, $(-i)^{\Nsat}$ and
$(\sigma^{z}_{c})^{\Nsat}$. We find that the resulting dynamics depends on $\Nsat$ only through its parity [see Fig.~\ref{fig:clifford}(a1)].

Following our analysis of the TQC regime, it is instructive to track the evolution of the system at Fibonacci times. We note that the Fibonacci construction $U_{n}=U_{n-2}U_{n-1}$ leads to the same recursion for the permutations discussed above. Thus, the Fibonacci-time oscillation periods can be obtained composing these permutations, with no reference to the quantum state. As detailed in the Supplemental Material~\cite{suppmat}, the Pauli labels give a period of $6$ in the Fibonacci index, for both parities, while the generator signs multiplies this by a further factor of $3$ for even $\Nsat$ and $2$ for odd $\Nsat$, giving oscillation periods of $18$ and $12$ respectively, as shown in Fig.~\ref{fig:clifford}(a2). The Fourier spectrum of the magnetization oscillations for even and odd values of $N_{\rm sat}$ differs substantially from each other, see Figs.~\ref{fig:clifford} (a3,a4). 
The above discussed Clifford structure governs the entanglement, since the central-spin reduced density matrix is pure only when $\mathcal{G}_{c}$ is supported on the central spin alone. The entanglement is therefore controlled by the label sector, which is insensitive to the generator signs, and its oscillation period in the Fibonacci index is $6$ for both parities. The entanglement dynamics at Fibonacci times is thus $6$ for all values of $N_{\rm sat}$, as seen in Figs.~\ref{fig:clifford} (a5, a6). The analysis so far has focused on the exact dynamics at the Clifford points. We have analyzed the robustness of the quasiperiodic oscillations away from the Clifford points by computing the TQC fraction, $f_{\rm TQC}$, corresponding to the quasiperiodic oscillations at the Clifford points for even and odd $N_{\rm sat}$, see Fig.~\ref{fig:clifford} (b). Interestingly, we find that a long-lived TQC regime characterized by a rigid quasiperiodic response, appears when $N_{\rm sat}$ is even.

\begin{figure}
\includegraphics[width=\linewidth]{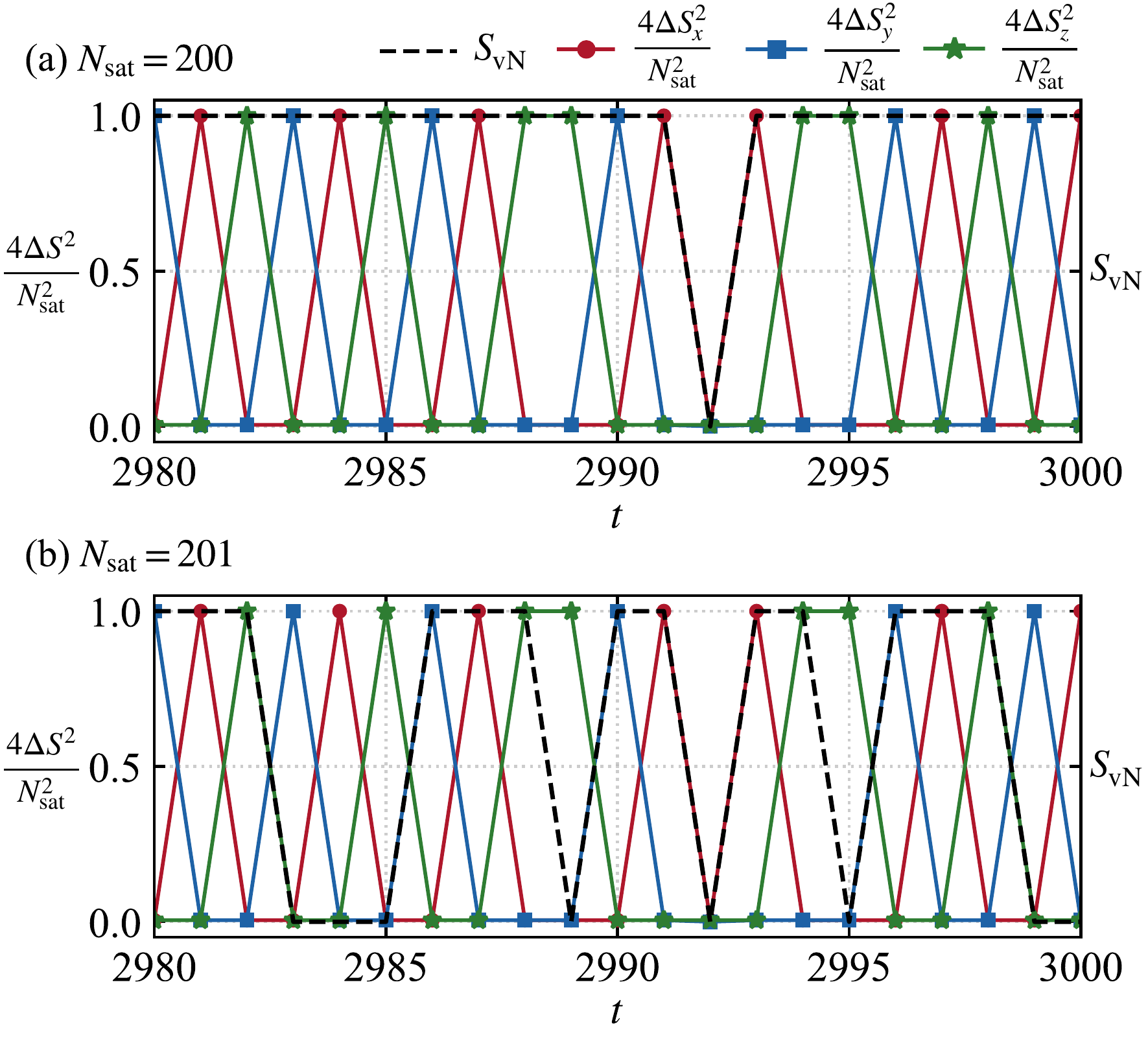}    
\caption{{\bf Entanglement dynamics at the Clifford points:} Time evolution of the entanglement entropy $\langle S_{\mathrm{vN}} \rangle$ (black dashed line) and the variances of the satellite spin magnetization at $4 \Delta S_{x,y,z}^2/N_{\rm sat}^2$ (red circles, blue squares and green stars) at the Clifford point $\theta =\pi/2, \lambda= \pi$ for $N_{\mathrm{sat}} = 200$ and $201$. For even $N_{\rm sat}$, the variance, $4 \Delta S_{\alpha}^2/N_{\rm sat}^2 = 1$, whenever $S_{\rm vN} = 1$, indicating the formation of Bell-cat states. For odd $N_{\rm sat}$, $4 \Delta S_{\alpha}^2/N_{\rm sat}^2 = 1$ for both $S_{\rm vN} = 1$ and $0$, signalling the generation of both Bell-cat and spin-cat states.}
\label{fig:entanglement}
\end{figure}

Before concluding this analysis, we note that the entanglement dynamics at the Clifford points exhibits parity dependence at stroboscopic times, with striking metrological implications [see Fig.~\ref{fig:entanglement} (a,b)]. Combining $S_{\rm vN}$ with the normalized collective variances $4 \Delta S_{\alpha}^2$ (where $S_{\alpha} = \sum_{i=1}^{N_{\rm sat}} S_i^{\alpha}$) of the satellite spins determine the form of the wavefunction at stroboscopic times. A variance saturating along a single axis $\alpha$ forces the bath to be supported with equal weight on the two extremal $\alpha$-eigenstates. The
entropy then determines whether that superposition is shared with the
central spin. When $S_{\mathrm{vN}}= 1$ the two branches are locked
to orthogonal central-spin states and the system is in a Bell-cat state~\cite{Lajcitopologically2026}. Conversely, when $S_{\mathrm{vN}}=0$ the central spin factorizes out entirely and the bath is left in a pure cat state of its own, $\big(|{+}\Nsat/2\rangle_{\alpha}
+e^{i\varphi}|{-}\Nsat/2\rangle_{\alpha}\big)/\sqrt{2}$, a spin-cat
state~\cite{PhysRevA.87.052323}. Neither diagnostic alone distinguishes these: the entropy does not see the polarization axis, and the variance does
not see whether the cat is shared. Crucially, we note that for pure states, the quantum Fisher information (QFI), $F_Q$, for a generator $S_{\rm \alpha}$ is given by $4 \Delta S_{\alpha}^2$ and it exhibits Heisenberg scaling $(F_Q \propto N_{\rm sat}^2)$, as expected from the Bell-cat and spin-cat states generated here. Our protocol thus naturally generates entangled states that can serve as resources for Heisenberg-limited metrology~\cite{giovannetti2006quantum,giovannetti2011advances}. Furthermore, since the collective variances are maximized at different times for different $\alpha$, this protocol can be employed to generate resource states for sensing along the $x$, $y$, and $z$ directions by controlling the number of kicks. Consequently, combining different runs with this protocol can be employed for multiparameter metrology.
 
\emph{Summary and Outlook:} In this work, we have examined the dynamics of a Fibonacci driven central spin model without disorder to investigate the emergence of TQC phases. We have analytically obtained the exact conditions, under which no entanglement is generated during the time evolution of the system, and demonstrated how this can lead to a robust $\pi$-pulse-proximate TQC regime, characterized by period-3 oscillations in Fibonacci time. Notably, while the exact quasiperiodicity of the oscillations is dictated by the Fibonacci drive, the non-generation of entanglement in these regimes only depends on the nature of the unitaries. Consequently, at these parameter regimes, no entanglement will be generated at stroboscopic times for other periodic or aperiodic drive protocols composed of these unitaries. Furthermore, we have analyzed the dynamics of the system at the Clifford points, and analytically characterized the magnetization and entanglement oscillations there. The quasiperiodic oscillations at these points are determined by the parity of $N_{\rm sat}$, and the quasiperiodic response persists even away from the Clifford points, when $N_{\rm sat}$ is even, leading to the formation of a Clifford-proximate TQC. Finally, we have demonstrated that the dynamical evolution at these Clifford points lead to the formation of Bell-cat and spin-cat states, which can be harnessed for multiparameter metrology. We note that a recent experiment has implemented Clifford gates in the central spin model in a NMR quantum processor~\cite{asthana2026efficientmanybodyshadowmetrology}. Our predictions can thus be verified in state-of-the-art experiments.

There are several interesting directions for future work. A natural first step would be to characterize the TQCs that emerge in a spin-$s$ central spin model~\cite{biswas2025discretetimecrystalsspins}. Furthermore, it would be interesting to study the dynamics of this system under other forms of aperiodic driving~\cite{PhysRevLett.126.040601,mori2021rigorous,PhysRevE.110.064150}. Finally, it would be interesting to investigate the fate of the TQCs when interactions are introduced amongst the satellite spins~\cite{PhysRevB.103.134201}.\\

\begin{acknowledgments}
S.C. acknowledges support from the Department of Science and Technology (DST), India, through Project No. DST/FFT/NQM/QSM/2024/3. T.N.  acknowledges the Advanced Research Grant (ARG) from
Anusandhan National Research Foundation Grant No.
ANRF/ARG/2025/003163/PS.
Generative AI tools (Anthropic's Claude) were used for editorial polishing and assistance with algebraic derivations. All scientific content, analyses, interpretations, and conclusions were developed and verified by the authors, who take full responsibility for the manuscript. 
\end{acknowledgments}
\bibliography{ref}

\begin{thebibliography}{77}%
\makeatletter
\providecommand \@ifxundefined [1]{%
 \@ifx{#1\undefined}
}%
\providecommand \@ifnum [1]{%
 \ifnum #1\expandafter \@firstoftwo
 \else \expandafter \@secondoftwo
 \fi
}%
\providecommand \@ifx [1]{%
 \ifx #1\expandafter \@firstoftwo
 \else \expandafter \@secondoftwo
 \fi
}%
\providecommand \natexlab [1]{#1}%
\providecommand \enquote  [1]{``#1''}%
\providecommand \bibnamefont  [1]{#1}%
\providecommand \bibfnamefont [1]{#1}%
\providecommand \citenamefont [1]{#1}%
\providecommand \href@noop [0]{\@secondoftwo}%
\providecommand \href [0]{\begingroup \@sanitize@url \@href}%
\providecommand \@href[1]{\@@startlink{#1}\@@href}%
\providecommand \@@href[1]{\endgroup#1\@@endlink}%
\providecommand \@sanitize@url [0]{\catcode `\\12\catcode `\$12\catcode
  `\&12\catcode `\#12\catcode `\^12\catcode `\_12\catcode `\%12\relax}%
\providecommand \@@startlink[1]{}%
\providecommand \@@endlink[0]{}%
\providecommand \url  [0]{\begingroup\@sanitize@url \@url }%
\providecommand \@url [1]{\endgroup\@href {#1}{\urlprefix }}%
\providecommand \urlprefix  [0]{URL }%
\providecommand \Eprint [0]{\href }%
\providecommand \doibase [0]{https://doi.org/}%
\providecommand \selectlanguage [0]{\@gobble}%
\providecommand \bibinfo  [0]{\@secondoftwo}%
\providecommand \bibfield  [0]{\@secondoftwo}%
\providecommand \translation [1]{[#1]}%
\providecommand \BibitemOpen [0]{}%
\providecommand \bibitemStop [0]{}%
\providecommand \bibitemNoStop [0]{.\EOS\space}%
\providecommand \EOS [0]{\spacefactor3000\relax}%
\providecommand \BibitemShut  [1]{\csname bibitem#1\endcsname}%
\let\auto@bib@innerbib\@empty
\bibitem [{\citenamefont {Bukov}\ \emph {et~al.}(2015)\citenamefont {Bukov},
  \citenamefont {D'Alessio},\ and\ \citenamefont
  {Polkovnikov}}]{bukov2015universal}%
  \BibitemOpen
  \bibfield  {author} {\bibinfo {author} {\bibfnamefont {M.}~\bibnamefont
  {Bukov}}, \bibinfo {author} {\bibfnamefont {L.}~\bibnamefont {D'Alessio}},\
  and\ \bibinfo {author} {\bibfnamefont {A.}~\bibnamefont {Polkovnikov}},\
  }\bibfield  {title} {\bibinfo {title} {Universal high-frequency behavior of
  periodically driven systems: from dynamical stabilization to floquet
  engineering},\ }\href@noop {} {\bibfield  {journal} {\bibinfo  {journal}
  {Advances in Physics}\ }\textbf {\bibinfo {volume} {64}},\ \bibinfo {pages}
  {139} (\bibinfo {year} {2015})}\BibitemShut {NoStop}%
\bibitem [{\citenamefont {Goldman}\ and\ \citenamefont
  {Dalibard}(2014)}]{goldman2014periodically}%
  \BibitemOpen
  \bibfield  {author} {\bibinfo {author} {\bibfnamefont {N.}~\bibnamefont
  {Goldman}}\ and\ \bibinfo {author} {\bibfnamefont {J.}~\bibnamefont
  {Dalibard}},\ }\bibfield  {title} {\bibinfo {title} {Periodically driven
  quantum systems: Effective hamiltonians and engineered gauge fields},\
  }\href@noop {} {\bibfield  {journal} {\bibinfo  {journal} {Physical review
  X}\ }\textbf {\bibinfo {volume} {4}},\ \bibinfo {pages} {031027} (\bibinfo
  {year} {2014})}\BibitemShut {NoStop}%
\bibitem [{\citenamefont {Oka}\ and\ \citenamefont
  {Kitamura}(2019)}]{oka2019floquet}%
  \BibitemOpen
  \bibfield  {author} {\bibinfo {author} {\bibfnamefont {T.}~\bibnamefont
  {Oka}}\ and\ \bibinfo {author} {\bibfnamefont {S.}~\bibnamefont {Kitamura}},\
  }\bibfield  {title} {\bibinfo {title} {Floquet engineering of quantum
  materials},\ }\href@noop {} {\bibfield  {journal} {\bibinfo  {journal}
  {Annual Review of Condensed Matter Physics}\ }\textbf {\bibinfo {volume}
  {10}},\ \bibinfo {pages} {387} (\bibinfo {year} {2019})}\BibitemShut
  {NoStop}%
\bibitem [{\citenamefont {Sen}\ \emph {et~al.}(2021)\citenamefont {Sen},
  \citenamefont {Sen},\ and\ \citenamefont {Sengupta}}]{sen2021analytic}%
  \BibitemOpen
  \bibfield  {author} {\bibinfo {author} {\bibfnamefont {A.}~\bibnamefont
  {Sen}}, \bibinfo {author} {\bibfnamefont {D.}~\bibnamefont {Sen}},\ and\
  \bibinfo {author} {\bibfnamefont {K.}~\bibnamefont {Sengupta}},\ }\bibfield
  {title} {\bibinfo {title} {Analytic approaches to periodically driven closed
  quantum systems: methods and applications},\ }\href@noop {} {\bibfield
  {journal} {\bibinfo  {journal} {Journal of Physics: Condensed Matter}\
  }\textbf {\bibinfo {volume} {33}},\ \bibinfo {pages} {443003} (\bibinfo
  {year} {2021})}\BibitemShut {NoStop}%
\bibitem [{\citenamefont {Weitenberg}\ and\ \citenamefont
  {Simonet}(2021)}]{Weitenberg2021}%
  \BibitemOpen
  \bibfield  {author} {\bibinfo {author} {\bibfnamefont {C.}~\bibnamefont
  {Weitenberg}}\ and\ \bibinfo {author} {\bibfnamefont {J.}~\bibnamefont
  {Simonet}},\ }\bibfield  {title} {\bibinfo {title} {Tailoring quantum gases
  by floquet engineering},\ }\href {https://doi.org/10.1038/s41567-021-01316-x}
  {\bibfield  {journal} {\bibinfo  {journal} {Nature Physics}\ }\textbf
  {\bibinfo {volume} {17}},\ \bibinfo {pages} {1342} (\bibinfo {year}
  {2021})}\BibitemShut {NoStop}%
\bibitem [{\citenamefont {Zhan}\ \emph {et~al.}(2024)\citenamefont {Zhan},
  \citenamefont {Chen}, \citenamefont {Ning}, \citenamefont {Ma}, \citenamefont
  {Wang}, \citenamefont {Xu},\ and\ \citenamefont
  {Wang}}]{zhan2024perspective}%
  \BibitemOpen
  \bibfield  {author} {\bibinfo {author} {\bibfnamefont {F.}~\bibnamefont
  {Zhan}}, \bibinfo {author} {\bibfnamefont {R.}~\bibnamefont {Chen}}, \bibinfo
  {author} {\bibfnamefont {Z.}~\bibnamefont {Ning}}, \bibinfo {author}
  {\bibfnamefont {D.-S.}\ \bibnamefont {Ma}}, \bibinfo {author} {\bibfnamefont
  {Z.}~\bibnamefont {Wang}}, \bibinfo {author} {\bibfnamefont {D.-H.}\
  \bibnamefont {Xu}},\ and\ \bibinfo {author} {\bibfnamefont {R.}~\bibnamefont
  {Wang}},\ }\bibfield  {title} {\bibinfo {title} {Perspective: Floquet
  engineering topological states from effective models towards realistic
  materials},\ }\href@noop {} {\bibfield  {journal} {\bibinfo  {journal}
  {Quantum Frontiers}\ }\textbf {\bibinfo {volume} {3}},\ \bibinfo {pages} {21}
  (\bibinfo {year} {2024})}\BibitemShut {NoStop}%
\bibitem [{\citenamefont {Rudner}\ and\ \citenamefont
  {Lindner}(2020)}]{rudner2020band}%
  \BibitemOpen
  \bibfield  {author} {\bibinfo {author} {\bibfnamefont {M.~S.}\ \bibnamefont
  {Rudner}}\ and\ \bibinfo {author} {\bibfnamefont {N.~H.}\ \bibnamefont
  {Lindner}},\ }\bibfield  {title} {\bibinfo {title} {Band structure
  engineering and non-equilibrium dynamics in floquet topological insulators},\
  }\href@noop {} {\bibfield  {journal} {\bibinfo  {journal} {Nature reviews
  physics}\ }\textbf {\bibinfo {volume} {2}},\ \bibinfo {pages} {229} (\bibinfo
  {year} {2020})}\BibitemShut {NoStop}%
\bibitem [{\citenamefont {Sato}\ and\ \citenamefont
  {Ikeda}(2025)}]{sato2025floquet}%
  \BibitemOpen
  \bibfield  {author} {\bibinfo {author} {\bibfnamefont {M.}~\bibnamefont
  {Sato}}\ and\ \bibinfo {author} {\bibfnamefont {T.~N.}\ \bibnamefont
  {Ikeda}},\ }\bibfield  {title} {\bibinfo {title} {Floquet theory and
  applications in open quantum and classical systems},\ }\href@noop {}
  {\bibfield  {journal} {\bibinfo  {journal} {Journal of the Physical Society
  of Japan}\ }\textbf {\bibinfo {volume} {94}},\ \bibinfo {pages} {111007}
  (\bibinfo {year} {2025})}\BibitemShut {NoStop}%
\bibitem [{\citenamefont {Sacha}\ and\ \citenamefont
  {Zakrzewski}(2018)}]{sacha2018time}%
  \BibitemOpen
  \bibfield  {author} {\bibinfo {author} {\bibfnamefont {K.}~\bibnamefont
  {Sacha}}\ and\ \bibinfo {author} {\bibfnamefont {J.}~\bibnamefont
  {Zakrzewski}},\ }\bibfield  {title} {\bibinfo {title} {Time crystals: a
  review},\ }\href@noop {} {\bibfield  {journal} {\bibinfo  {journal} {Reports
  on Progress in Physics}\ }\textbf {\bibinfo {volume} {81}},\ \bibinfo {pages}
  {016401} (\bibinfo {year} {2018})}\BibitemShut {NoStop}%
\bibitem [{\citenamefont {Else}\ \emph
  {et~al.}(2020{\natexlab{a}})\citenamefont {Else}, \citenamefont {Monroe},
  \citenamefont {Nayak},\ and\ \citenamefont {Yao}}]{else2020discrete}%
  \BibitemOpen
  \bibfield  {author} {\bibinfo {author} {\bibfnamefont {D.~V.}\ \bibnamefont
  {Else}}, \bibinfo {author} {\bibfnamefont {C.}~\bibnamefont {Monroe}},
  \bibinfo {author} {\bibfnamefont {C.}~\bibnamefont {Nayak}},\ and\ \bibinfo
  {author} {\bibfnamefont {N.~Y.}\ \bibnamefont {Yao}},\ }\bibfield  {title}
  {\bibinfo {title} {Discrete time crystals},\ }\href@noop {} {\bibfield
  {journal} {\bibinfo  {journal} {Annual Review of Condensed Matter Physics}\
  }\textbf {\bibinfo {volume} {11}},\ \bibinfo {pages} {467} (\bibinfo {year}
  {2020}{\natexlab{a}})}\BibitemShut {NoStop}%
\bibitem [{\citenamefont {Khemani}\ \emph {et~al.}(2019)\citenamefont
  {Khemani}, \citenamefont {Moessner},\ and\ \citenamefont
  {Sondhi}}]{khemani2019brief}%
  \BibitemOpen
  \bibfield  {author} {\bibinfo {author} {\bibfnamefont {V.}~\bibnamefont
  {Khemani}}, \bibinfo {author} {\bibfnamefont {R.}~\bibnamefont {Moessner}},\
  and\ \bibinfo {author} {\bibfnamefont {S.}~\bibnamefont {Sondhi}},\
  }\bibfield  {title} {\bibinfo {title} {A brief history of time crystals},\
  }\href@noop {} {\bibfield  {journal} {\bibinfo  {journal} {arXiv preprint
  arXiv:1910.10745}\ } (\bibinfo {year} {2019})}\BibitemShut {NoStop}%
\bibitem [{\citenamefont {Zaletel}\ \emph {et~al.}(2023)\citenamefont
  {Zaletel}, \citenamefont {Lukin}, \citenamefont {Monroe}, \citenamefont
  {Nayak}, \citenamefont {Wilczek},\ and\ \citenamefont
  {Yao}}]{zaletel2023colloquium}%
  \BibitemOpen
  \bibfield  {author} {\bibinfo {author} {\bibfnamefont {M.~P.}\ \bibnamefont
  {Zaletel}}, \bibinfo {author} {\bibfnamefont {M.}~\bibnamefont {Lukin}},
  \bibinfo {author} {\bibfnamefont {C.}~\bibnamefont {Monroe}}, \bibinfo
  {author} {\bibfnamefont {C.}~\bibnamefont {Nayak}}, \bibinfo {author}
  {\bibfnamefont {F.}~\bibnamefont {Wilczek}},\ and\ \bibinfo {author}
  {\bibfnamefont {N.~Y.}\ \bibnamefont {Yao}},\ }\bibfield  {title} {\bibinfo
  {title} {Colloquium: Quantum and classical discrete time crystals},\
  }\href@noop {} {\bibfield  {journal} {\bibinfo  {journal} {Reviews of Modern
  Physics}\ }\textbf {\bibinfo {volume} {95}},\ \bibinfo {pages} {031001}
  (\bibinfo {year} {2023})}\BibitemShut {NoStop}%
\bibitem [{\citenamefont {Rudner}\ \emph {et~al.}(2013)\citenamefont {Rudner},
  \citenamefont {Lindner}, \citenamefont {Berg},\ and\ \citenamefont
  {Levin}}]{rudner2013anomalous}%
  \BibitemOpen
  \bibfield  {author} {\bibinfo {author} {\bibfnamefont {M.~S.}\ \bibnamefont
  {Rudner}}, \bibinfo {author} {\bibfnamefont {N.~H.}\ \bibnamefont {Lindner}},
  \bibinfo {author} {\bibfnamefont {E.}~\bibnamefont {Berg}},\ and\ \bibinfo
  {author} {\bibfnamefont {M.}~\bibnamefont {Levin}},\ }\bibfield  {title}
  {\bibinfo {title} {Anomalous edge states and the bulk-edge correspondence for
  periodically driven two-dimensional systems},\ }\href@noop {} {\bibfield
  {journal} {\bibinfo  {journal} {Physical Review X}\ }\textbf {\bibinfo
  {volume} {3}},\ \bibinfo {pages} {031005} (\bibinfo {year}
  {2013})}\BibitemShut {NoStop}%
\bibitem [{\citenamefont {Nathan}\ \emph {et~al.}(2019)\citenamefont {Nathan},
  \citenamefont {Abanin}, \citenamefont {Berg}, \citenamefont {Lindner},\ and\
  \citenamefont {Rudner}}]{nathan2019anomalous}%
  \BibitemOpen
  \bibfield  {author} {\bibinfo {author} {\bibfnamefont {F.}~\bibnamefont
  {Nathan}}, \bibinfo {author} {\bibfnamefont {D.}~\bibnamefont {Abanin}},
  \bibinfo {author} {\bibfnamefont {E.}~\bibnamefont {Berg}}, \bibinfo {author}
  {\bibfnamefont {N.~H.}\ \bibnamefont {Lindner}},\ and\ \bibinfo {author}
  {\bibfnamefont {M.~S.}\ \bibnamefont {Rudner}},\ }\bibfield  {title}
  {\bibinfo {title} {Anomalous floquet insulators},\ }\href@noop {} {\bibfield
  {journal} {\bibinfo  {journal} {Physical Review B}\ }\textbf {\bibinfo
  {volume} {99}},\ \bibinfo {pages} {195133} (\bibinfo {year}
  {2019})}\BibitemShut {NoStop}%
\bibitem [{\citenamefont {Qiu}\ \emph {et~al.}(2026)\citenamefont {Qiu},
  \citenamefont {Tong}, \citenamefont {Zhang}, \citenamefont {Zhang},\ and\
  \citenamefont {Qiu}}]{qiu2026observation}%
  \BibitemOpen
  \bibfield  {author} {\bibinfo {author} {\bibfnamefont {H.}~\bibnamefont
  {Qiu}}, \bibinfo {author} {\bibfnamefont {S.}~\bibnamefont {Tong}}, \bibinfo
  {author} {\bibfnamefont {Q.}~\bibnamefont {Zhang}}, \bibinfo {author}
  {\bibfnamefont {K.}~\bibnamefont {Zhang}},\ and\ \bibinfo {author}
  {\bibfnamefont {C.}~\bibnamefont {Qiu}},\ }\bibfield  {title} {\bibinfo
  {title} {Observation of anomalous floquet non-abelian topological
  insulators},\ }\href@noop {} {\bibfield  {journal} {\bibinfo  {journal}
  {Physical Review X}\ }\textbf {\bibinfo {volume} {16}},\ \bibinfo {pages}
  {011061} (\bibinfo {year} {2026})}\BibitemShut {NoStop}%
\bibitem [{\citenamefont {Li}\ and\ \citenamefont {Hu}(2023)}]{li2023floquet}%
  \BibitemOpen
  \bibfield  {author} {\bibinfo {author} {\bibfnamefont {T.}~\bibnamefont
  {Li}}\ and\ \bibinfo {author} {\bibfnamefont {H.}~\bibnamefont {Hu}},\
  }\bibfield  {title} {\bibinfo {title} {Floquet non-abelian topological
  insulator and multifold bulk-edge correspondence},\ }\href@noop {} {\bibfield
   {journal} {\bibinfo  {journal} {Nature Communications}\ }\textbf {\bibinfo
  {volume} {14}},\ \bibinfo {pages} {6418} (\bibinfo {year}
  {2023})}\BibitemShut {NoStop}%
\bibitem [{\citenamefont {Nag}\ and\ \citenamefont
  {Roy}(2021)}]{nag2021anomalous}%
  \BibitemOpen
  \bibfield  {author} {\bibinfo {author} {\bibfnamefont {T.}~\bibnamefont
  {Nag}}\ and\ \bibinfo {author} {\bibfnamefont {B.}~\bibnamefont {Roy}},\
  }\bibfield  {title} {\bibinfo {title} {Anomalous and normal dislocation modes
  in floquet topological insulators},\ }\href@noop {} {\bibfield  {journal}
  {\bibinfo  {journal} {Communications Physics}\ }\textbf {\bibinfo {volume}
  {4}},\ \bibinfo {pages} {157} (\bibinfo {year} {2021})}\BibitemShut {NoStop}%
\bibitem [{\citenamefont {Ghosh}\ \emph {et~al.}(2022)\citenamefont {Ghosh},
  \citenamefont {Nag},\ and\ \citenamefont {Saha}}]{PhysRevB.105.155406}%
  \BibitemOpen
  \bibfield  {author} {\bibinfo {author} {\bibfnamefont {A.~K.}\ \bibnamefont
  {Ghosh}}, \bibinfo {author} {\bibfnamefont {T.}~\bibnamefont {Nag}},\ and\
  \bibinfo {author} {\bibfnamefont {A.}~\bibnamefont {Saha}},\ }\bibfield
  {title} {\bibinfo {title} {Dynamical construction of quadrupolar and
  octupolar topological superconductors},\ }\href
  {https://doi.org/10.1103/PhysRevB.105.155406} {\bibfield  {journal} {\bibinfo
   {journal} {Phys. Rev. B}\ }\textbf {\bibinfo {volume} {105}},\ \bibinfo
  {pages} {155406} (\bibinfo {year} {2022})}\BibitemShut {NoStop}%
\bibitem [{\citenamefont {Ghosh}\ \emph {et~al.}(2024)\citenamefont {Ghosh},
  \citenamefont {Nag},\ and\ \citenamefont {Saha}}]{ghosh2024generation}%
  \BibitemOpen
  \bibfield  {author} {\bibinfo {author} {\bibfnamefont {A.~K.}\ \bibnamefont
  {Ghosh}}, \bibinfo {author} {\bibfnamefont {T.}~\bibnamefont {Nag}},\ and\
  \bibinfo {author} {\bibfnamefont {A.}~\bibnamefont {Saha}},\ }\bibfield
  {title} {\bibinfo {title} {Generation of higher-order topological insulators
  using periodic driving},\ }\href@noop {} {\bibfield  {journal} {\bibinfo
  {journal} {Journal of Physics: Condensed Matter}\ }\textbf {\bibinfo {volume}
  {36}},\ \bibinfo {pages} {093001} (\bibinfo {year} {2024})}\BibitemShut
  {NoStop}%
\bibitem [{\citenamefont {Pal}\ \emph {et~al.}(2025)\citenamefont {Pal},
  \citenamefont {Mondal}, \citenamefont {Nag},\ and\ \citenamefont
  {Saha}}]{prnx-47mk}%
  \BibitemOpen
  \bibfield  {author} {\bibinfo {author} {\bibfnamefont {A.}~\bibnamefont
  {Pal}}, \bibinfo {author} {\bibfnamefont {D.}~\bibnamefont {Mondal}},
  \bibinfo {author} {\bibfnamefont {T.}~\bibnamefont {Nag}},\ and\ \bibinfo
  {author} {\bibfnamefont {A.}~\bibnamefont {Saha}},\ }\bibfield  {title}
  {\bibinfo {title} {Josephson current signature of floquet majorana and
  topological accidental zero modes in altermagnet heterostructures},\ }\href
  {https://doi.org/10.1103/prnx-47mk} {\bibfield  {journal} {\bibinfo
  {journal} {Phys. Rev. B}\ }\textbf {\bibinfo {volume} {112}},\ \bibinfo
  {pages} {L201408} (\bibinfo {year} {2025})}\BibitemShut {NoStop}%
\bibitem [{\citenamefont {Claeys}\ \emph {et~al.}(2019)\citenamefont {Claeys},
  \citenamefont {Pandey}, \citenamefont {Sels},\ and\ \citenamefont
  {Polkovnikov}}]{PhysRevLett.123.090602}%
  \BibitemOpen
  \bibfield  {author} {\bibinfo {author} {\bibfnamefont {P.~W.}\ \bibnamefont
  {Claeys}}, \bibinfo {author} {\bibfnamefont {M.}~\bibnamefont {Pandey}},
  \bibinfo {author} {\bibfnamefont {D.}~\bibnamefont {Sels}},\ and\ \bibinfo
  {author} {\bibfnamefont {A.}~\bibnamefont {Polkovnikov}},\ }\bibfield
  {title} {\bibinfo {title} {Floquet-engineering counterdiabatic protocols in
  quantum many-body systems},\ }\href
  {https://doi.org/10.1103/PhysRevLett.123.090602} {\bibfield  {journal}
  {\bibinfo  {journal} {Phys. Rev. Lett.}\ }\textbf {\bibinfo {volume} {123}},\
  \bibinfo {pages} {090602} (\bibinfo {year} {2019})}\BibitemShut {NoStop}%
\bibitem [{\citenamefont {Ritter}\ \emph {et~al.}(2025)\citenamefont {Ritter},
  \citenamefont {Long}, \citenamefont {Yue}, \citenamefont {Chandran},\ and\
  \citenamefont {Koll\'ar}}]{3dpp-p2pr}%
  \BibitemOpen
  \bibfield  {author} {\bibinfo {author} {\bibfnamefont {M.}~\bibnamefont
  {Ritter}}, \bibinfo {author} {\bibfnamefont {D.~M.}\ \bibnamefont {Long}},
  \bibinfo {author} {\bibfnamefont {Q.}~\bibnamefont {Yue}}, \bibinfo {author}
  {\bibfnamefont {A.}~\bibnamefont {Chandran}},\ and\ \bibinfo {author}
  {\bibfnamefont {A.~J.}\ \bibnamefont {Koll\'ar}},\ }\bibfield  {title}
  {\bibinfo {title} {Autonomous stabilization of floquet states using static
  dissipation},\ }\href {https://doi.org/10.1103/3dpp-p2pr} {\bibfield
  {journal} {\bibinfo  {journal} {Phys. Rev. X}\ }\textbf {\bibinfo {volume}
  {15}},\ \bibinfo {pages} {031028} (\bibinfo {year} {2025})}\BibitemShut
  {NoStop}%
\bibitem [{\citenamefont {Gangopadhay}\ and\ \citenamefont
  {Choudhury}(2025)}]{Gangopadhay_2025}%
  \BibitemOpen
  \bibfield  {author} {\bibinfo {author} {\bibfnamefont {N.}~\bibnamefont
  {Gangopadhay}}\ and\ \bibinfo {author} {\bibfnamefont {S.}~\bibnamefont
  {Choudhury}},\ }\bibfield  {title} {\bibinfo {title} {Counterdiabatic route
  to entanglement steering and dynamical freezing in the floquet
  lipkin-meshkov-glick model},\ }\href {http://dx.doi.org/10.1103/bzcf-gm89}
  {\bibfield  {journal} {\bibinfo  {journal} {Physical Review Letters}\
  }\textbf {\bibinfo {volume} {135}} (\bibinfo {year} {2025})}\BibitemShut
  {NoStop}%
\bibitem [{\citenamefont {Tamang}\ \emph {et~al.}(2021)\citenamefont {Tamang},
  \citenamefont {Nag},\ and\ \citenamefont {Biswas}}]{PhysRevB.104.174308}%
  \BibitemOpen
  \bibfield  {author} {\bibinfo {author} {\bibfnamefont {L.}~\bibnamefont
  {Tamang}}, \bibinfo {author} {\bibfnamefont {T.}~\bibnamefont {Nag}},\ and\
  \bibinfo {author} {\bibfnamefont {T.}~\bibnamefont {Biswas}},\ }\bibfield
  {title} {\bibinfo {title} {Floquet engineering of low-energy dispersions and
  dynamical localization in a periodically kicked three-band system},\ }\href
  {https://doi.org/10.1103/PhysRevB.104.174308} {\bibfield  {journal} {\bibinfo
   {journal} {Phys. Rev. B}\ }\textbf {\bibinfo {volume} {104}},\ \bibinfo
  {pages} {174308} (\bibinfo {year} {2021})}\BibitemShut {NoStop}%
\bibitem [{\citenamefont {Nag}\ \emph {et~al.}(2019)\citenamefont {Nag},
  \citenamefont {Slager}, \citenamefont {Higuchi},\ and\ \citenamefont
  {Oka}}]{PhysRevB.100.134301}%
  \BibitemOpen
  \bibfield  {author} {\bibinfo {author} {\bibfnamefont {T.}~\bibnamefont
  {Nag}}, \bibinfo {author} {\bibfnamefont {R.-J.}\ \bibnamefont {Slager}},
  \bibinfo {author} {\bibfnamefont {T.}~\bibnamefont {Higuchi}},\ and\ \bibinfo
  {author} {\bibfnamefont {T.}~\bibnamefont {Oka}},\ }\bibfield  {title}
  {\bibinfo {title} {Dynamical synchronization transition in interacting
  electron systems},\ }\href {https://doi.org/10.1103/PhysRevB.100.134301}
  {\bibfield  {journal} {\bibinfo  {journal} {Phys. Rev. B}\ }\textbf {\bibinfo
  {volume} {100}},\ \bibinfo {pages} {134301} (\bibinfo {year}
  {2019})}\BibitemShut {NoStop}%
\bibitem [{\citenamefont {Nag}\ \emph {et~al.}(2014)\citenamefont {Nag},
  \citenamefont {Roy}, \citenamefont {Dutta},\ and\ \citenamefont
  {Sen}}]{PhysRevB.89.165425}%
  \BibitemOpen
  \bibfield  {author} {\bibinfo {author} {\bibfnamefont {T.}~\bibnamefont
  {Nag}}, \bibinfo {author} {\bibfnamefont {S.}~\bibnamefont {Roy}}, \bibinfo
  {author} {\bibfnamefont {A.}~\bibnamefont {Dutta}},\ and\ \bibinfo {author}
  {\bibfnamefont {D.}~\bibnamefont {Sen}},\ }\bibfield  {title} {\bibinfo
  {title} {Dynamical localization in a chain of hard core bosons under periodic
  driving},\ }\href {https://doi.org/10.1103/PhysRevB.89.165425} {\bibfield
  {journal} {\bibinfo  {journal} {Phys. Rev. B}\ }\textbf {\bibinfo {volume}
  {89}},\ \bibinfo {pages} {165425} (\bibinfo {year} {2014})}\BibitemShut
  {NoStop}%
\bibitem [{\citenamefont {Iemini}\ \emph {et~al.}(2024)\citenamefont {Iemini},
  \citenamefont {Fazio},\ and\ \citenamefont {Sanpera}}]{PhysRevA.109.L050203}%
  \BibitemOpen
  \bibfield  {author} {\bibinfo {author} {\bibfnamefont {F.}~\bibnamefont
  {Iemini}}, \bibinfo {author} {\bibfnamefont {R.}~\bibnamefont {Fazio}},\ and\
  \bibinfo {author} {\bibfnamefont {A.}~\bibnamefont {Sanpera}},\ }\bibfield
  {title} {\bibinfo {title} {Floquet time crystals as quantum sensors of ac
  fields},\ }\href {https://doi.org/10.1103/PhysRevA.109.L050203} {\bibfield
  {journal} {\bibinfo  {journal} {Phys. Rev. A}\ }\textbf {\bibinfo {volume}
  {109}},\ \bibinfo {pages} {L050203} (\bibinfo {year} {2024})}\BibitemShut
  {NoStop}%
\bibitem [{\citenamefont {Shukla}\ \emph {et~al.}(2025)\citenamefont {Shukla},
  \citenamefont {Chotorlishvili}, \citenamefont {Mishra},\ and\ \citenamefont
  {Iemini}}]{PhysRevB.111.024315}%
  \BibitemOpen
  \bibfield  {author} {\bibinfo {author} {\bibfnamefont {R.~K.}\ \bibnamefont
  {Shukla}}, \bibinfo {author} {\bibfnamefont {L.}~\bibnamefont
  {Chotorlishvili}}, \bibinfo {author} {\bibfnamefont {S.~K.}\ \bibnamefont
  {Mishra}},\ and\ \bibinfo {author} {\bibfnamefont {F.}~\bibnamefont
  {Iemini}},\ }\bibfield  {title} {\bibinfo {title} {Prethermal floquet time
  crystals in chiral multiferroic chains and applications as quantum sensors of
  ac fields},\ }\href {https://doi.org/10.1103/PhysRevB.111.024315} {\bibfield
  {journal} {\bibinfo  {journal} {Phys. Rev. B}\ }\textbf {\bibinfo {volume}
  {111}},\ \bibinfo {pages} {024315} (\bibinfo {year} {2025})}\BibitemShut
  {NoStop}%
\bibitem [{\citenamefont {Sharma}\ \emph {et~al.}(2026)\citenamefont {Sharma},
  \citenamefont {Choudhury},\ and\ \citenamefont
  {Bandyopadhyay}}]{sharma2026quantummetrologypartiallyaccessible}%
  \BibitemOpen
  \bibfield  {author} {\bibinfo {author} {\bibfnamefont {H.}~\bibnamefont
  {Sharma}}, \bibinfo {author} {\bibfnamefont {S.}~\bibnamefont {Choudhury}},\
  and\ \bibinfo {author} {\bibfnamefont {J.~N.}\ \bibnamefont
  {Bandyopadhyay}},\ }\href {https://arxiv.org/abs/2602.12914} {\bibinfo
  {title} {Quantum metrology with partially accessible chaotic sensors}}
  (\bibinfo {year} {2026}),\ \Eprint {https://arxiv.org/abs/2602.12914}
  {arXiv:2602.12914 [quant-ph]} \BibitemShut {NoStop}%
\bibitem [{\citenamefont {Fiderer}\ and\ \citenamefont
  {Braun}(2018)}]{fiderer2018quantum}%
  \BibitemOpen
  \bibfield  {author} {\bibinfo {author} {\bibfnamefont {L.~J.}\ \bibnamefont
  {Fiderer}}\ and\ \bibinfo {author} {\bibfnamefont {D.}~\bibnamefont
  {Braun}},\ }\bibfield  {title} {\bibinfo {title} {Quantum metrology with
  quantum-chaotic sensors},\ }\href@noop {} {\bibfield  {journal} {\bibinfo
  {journal} {Nature communications}\ }\textbf {\bibinfo {volume} {9}},\
  \bibinfo {pages} {1351} (\bibinfo {year} {2018})}\BibitemShut {NoStop}%
\bibitem [{\citenamefont {Engelhardt}\ \emph {et~al.}(2024)\citenamefont
  {Engelhardt}, \citenamefont {Choudhury},\ and\ \citenamefont
  {Liu}}]{PhysRevResearch.6.013116}%
  \BibitemOpen
  \bibfield  {author} {\bibinfo {author} {\bibfnamefont {G.}~\bibnamefont
  {Engelhardt}}, \bibinfo {author} {\bibfnamefont {S.}~\bibnamefont
  {Choudhury}},\ and\ \bibinfo {author} {\bibfnamefont {W.~V.}\ \bibnamefont
  {Liu}},\ }\bibfield  {title} {\bibinfo {title} {Unified light-matter floquet
  theory and its application to quantum communication},\ }\href
  {https://doi.org/10.1103/PhysRevResearch.6.013116} {\bibfield  {journal}
  {\bibinfo  {journal} {Phys. Rev. Res.}\ }\textbf {\bibinfo {volume} {6}},\
  \bibinfo {pages} {013116} (\bibinfo {year} {2024})}\BibitemShut {NoStop}%
\bibitem [{\citenamefont {Geier}\ \emph {et~al.}(2021)\citenamefont {Geier},
  \citenamefont {Thaicharoen}, \citenamefont {Hainaut}, \citenamefont {Franz},
  \citenamefont {Salzinger}, \citenamefont {Tebben}, \citenamefont
  {Grimshandl}, \citenamefont {Z{\"u}rn},\ and\ \citenamefont
  {Weidem{\"u}ller}}]{geier2021floquet}%
  \BibitemOpen
  \bibfield  {author} {\bibinfo {author} {\bibfnamefont {S.}~\bibnamefont
  {Geier}}, \bibinfo {author} {\bibfnamefont {N.}~\bibnamefont {Thaicharoen}},
  \bibinfo {author} {\bibfnamefont {C.}~\bibnamefont {Hainaut}}, \bibinfo
  {author} {\bibfnamefont {T.}~\bibnamefont {Franz}}, \bibinfo {author}
  {\bibfnamefont {A.}~\bibnamefont {Salzinger}}, \bibinfo {author}
  {\bibfnamefont {A.}~\bibnamefont {Tebben}}, \bibinfo {author} {\bibfnamefont
  {D.}~\bibnamefont {Grimshandl}}, \bibinfo {author} {\bibfnamefont
  {G.}~\bibnamefont {Z{\"u}rn}},\ and\ \bibinfo {author} {\bibfnamefont
  {M.}~\bibnamefont {Weidem{\"u}ller}},\ }\bibfield  {title} {\bibinfo {title}
  {Floquet hamiltonian engineering of an isolated many-body spin system},\
  }\href@noop {} {\bibfield  {journal} {\bibinfo  {journal} {Science}\ }\textbf
  {\bibinfo {volume} {374}},\ \bibinfo {pages} {1149} (\bibinfo {year}
  {2021})}\BibitemShut {NoStop}%
\bibitem [{\citenamefont {Katz}\ \emph {et~al.}(2025)\citenamefont {Katz},
  \citenamefont {Feng}, \citenamefont {Porras},\ and\ \citenamefont
  {Monroe}}]{katz2025floquet}%
  \BibitemOpen
  \bibfield  {author} {\bibinfo {author} {\bibfnamefont {O.}~\bibnamefont
  {Katz}}, \bibinfo {author} {\bibfnamefont {L.}~\bibnamefont {Feng}}, \bibinfo
  {author} {\bibfnamefont {D.}~\bibnamefont {Porras}},\ and\ \bibinfo {author}
  {\bibfnamefont {C.}~\bibnamefont {Monroe}},\ }\bibfield  {title} {\bibinfo
  {title} {Floquet control of interactions and edge states in a programmable
  quantum simulator},\ }\href@noop {} {\bibfield  {journal} {\bibinfo
  {journal} {Nature Communications}\ }\textbf {\bibinfo {volume} {16}},\
  \bibinfo {pages} {8815} (\bibinfo {year} {2025})}\BibitemShut {NoStop}%
\bibitem [{\citenamefont {Verdeny}\ \emph {et~al.}(2016)\citenamefont
  {Verdeny}, \citenamefont {Puig},\ and\ \citenamefont
  {Mintert}}]{verdeny2016quasi}%
  \BibitemOpen
  \bibfield  {author} {\bibinfo {author} {\bibfnamefont {A.}~\bibnamefont
  {Verdeny}}, \bibinfo {author} {\bibfnamefont {J.}~\bibnamefont {Puig}},\ and\
  \bibinfo {author} {\bibfnamefont {F.}~\bibnamefont {Mintert}},\ }\bibfield
  {title} {\bibinfo {title} {Quasi-periodically driven quantum systems},\
  }\href@noop {} {\bibfield  {journal} {\bibinfo  {journal} {Zeitschrift
  f{\"u}r Naturforschung A}\ }\textbf {\bibinfo {volume} {71}},\ \bibinfo
  {pages} {897} (\bibinfo {year} {2016})}\BibitemShut {NoStop}%
\bibitem [{\citenamefont {Else}\ \emph
  {et~al.}(2020{\natexlab{b}})\citenamefont {Else}, \citenamefont {Ho},\ and\
  \citenamefont {Dumitrescu}}]{else2020long}%
  \BibitemOpen
  \bibfield  {author} {\bibinfo {author} {\bibfnamefont {D.~V.}\ \bibnamefont
  {Else}}, \bibinfo {author} {\bibfnamefont {W.~W.}\ \bibnamefont {Ho}},\ and\
  \bibinfo {author} {\bibfnamefont {P.~T.}\ \bibnamefont {Dumitrescu}},\
  }\bibfield  {title} {\bibinfo {title} {Long-lived interacting phases of
  matter protected by multiple time-translation symmetries in quasiperiodically
  driven systems},\ }\href@noop {} {\bibfield  {journal} {\bibinfo  {journal}
  {Physical Review X}\ }\textbf {\bibinfo {volume} {10}},\ \bibinfo {pages}
  {021032} (\bibinfo {year} {2020}{\natexlab{b}})}\BibitemShut {NoStop}%
\bibitem [{\citenamefont {Gallone}\ and\ \citenamefont
  {Langella}(2024)}]{gallone2024prethermalization}%
  \BibitemOpen
  \bibfield  {author} {\bibinfo {author} {\bibfnamefont {M.}~\bibnamefont
  {Gallone}}\ and\ \bibinfo {author} {\bibfnamefont {B.}~\bibnamefont
  {Langella}},\ }\bibfield  {title} {\bibinfo {title} {Prethermalization and
  conservation laws in quasi-periodically driven quantum systems},\ }\href@noop
  {} {\bibfield  {journal} {\bibinfo  {journal} {Journal of Statistical
  Physics}\ }\textbf {\bibinfo {volume} {191}},\ \bibinfo {pages} {100}
  (\bibinfo {year} {2024})}\BibitemShut {NoStop}%
\bibitem [{\citenamefont {Anisur}\ \emph {et~al.}(2026)\citenamefont {Anisur},
  \citenamefont {Mondal}, \citenamefont {Ray},\ and\ \citenamefont
  {Choudhury}}]{anisur2026dissipativedicketimequasicrystals}%
  \BibitemOpen
  \bibfield  {author} {\bibinfo {author} {\bibfnamefont {S.}~\bibnamefont
  {Anisur}}, \bibinfo {author} {\bibfnamefont {S.}~\bibnamefont {Mondal}},
  \bibinfo {author} {\bibfnamefont {T.}~\bibnamefont {Ray}},\ and\ \bibinfo
  {author} {\bibfnamefont {S.}~\bibnamefont {Choudhury}},\ }\href
  {https://arxiv.org/abs/2602.05994} {\bibinfo {title} {Dissipation stabilizes
  dicke time quasicrystals}} (\bibinfo {year} {2026}),\ \Eprint
  {https://arxiv.org/abs/2602.05994} {arXiv:2602.05994 [quant-ph]} \BibitemShut
  {NoStop}%
\bibitem [{\citenamefont {Anisur}\ \emph {et~al.}(2025)\citenamefont {Anisur},
  \citenamefont {Liu},\ and\ \citenamefont {Choudhury}}]{anisur2025quasi}%
  \BibitemOpen
  \bibfield  {author} {\bibinfo {author} {\bibfnamefont {S.}~\bibnamefont
  {Anisur}}, \bibinfo {author} {\bibfnamefont {W.~V.}\ \bibnamefont {Liu}},\
  and\ \bibinfo {author} {\bibfnamefont {S.}~\bibnamefont {Choudhury}},\
  }\bibfield  {title} {\bibinfo {title} {Quasi-discrete time crystals in the
  quasiperiodically driven lipkin--meshkov--glick model},\ }\href@noop {}
  {\bibfield  {journal} {\bibinfo  {journal} {Entropy}\ }\textbf {\bibinfo
  {volume} {27}},\ \bibinfo {pages} {609} (\bibinfo {year} {2025})}\BibitemShut
  {NoStop}%
\bibitem [{\citenamefont {Dumitrescu}\ \emph {et~al.}(2018)\citenamefont
  {Dumitrescu}, \citenamefont {Vasseur},\ and\ \citenamefont
  {Potter}}]{PhysRevLett.120.070602}%
  \BibitemOpen
  \bibfield  {author} {\bibinfo {author} {\bibfnamefont {P.~T.}\ \bibnamefont
  {Dumitrescu}}, \bibinfo {author} {\bibfnamefont {R.}~\bibnamefont
  {Vasseur}},\ and\ \bibinfo {author} {\bibfnamefont {A.~C.}\ \bibnamefont
  {Potter}},\ }\bibfield  {title} {\bibinfo {title} {Logarithmically slow
  relaxation in quasiperiodically driven random spin chains},\ }\href
  {https://doi.org/10.1103/PhysRevLett.120.070602} {\bibfield  {journal}
  {\bibinfo  {journal} {Phys. Rev. Lett.}\ }\textbf {\bibinfo {volume} {120}},\
  \bibinfo {pages} {070602} (\bibinfo {year} {2018})}\BibitemShut {NoStop}%
\bibitem [{\citenamefont {Dumitrescu}\ \emph {et~al.}(2022)\citenamefont
  {Dumitrescu}, \citenamefont {Bohnet}, \citenamefont {Gaebler}, \citenamefont
  {Hankin}, \citenamefont {Hayes}, \citenamefont {Kumar}, \citenamefont
  {Neyenhuis}, \citenamefont {Vasseur},\ and\ \citenamefont
  {Potter}}]{dumitrescu2022dynamical}%
  \BibitemOpen
  \bibfield  {author} {\bibinfo {author} {\bibfnamefont {P.~T.}\ \bibnamefont
  {Dumitrescu}}, \bibinfo {author} {\bibfnamefont {J.~G.}\ \bibnamefont
  {Bohnet}}, \bibinfo {author} {\bibfnamefont {J.~P.}\ \bibnamefont {Gaebler}},
  \bibinfo {author} {\bibfnamefont {A.}~\bibnamefont {Hankin}}, \bibinfo
  {author} {\bibfnamefont {D.}~\bibnamefont {Hayes}}, \bibinfo {author}
  {\bibfnamefont {A.}~\bibnamefont {Kumar}}, \bibinfo {author} {\bibfnamefont
  {B.}~\bibnamefont {Neyenhuis}}, \bibinfo {author} {\bibfnamefont
  {R.}~\bibnamefont {Vasseur}},\ and\ \bibinfo {author} {\bibfnamefont {A.~C.}\
  \bibnamefont {Potter}},\ }\bibfield  {title} {\bibinfo {title} {Dynamical
  topological phase realized in a trapped-ion quantum simulator},\ }\href@noop
  {} {\bibfield  {journal} {\bibinfo  {journal} {Nature}\ }\textbf {\bibinfo
  {volume} {607}},\ \bibinfo {pages} {463} (\bibinfo {year}
  {2022})}\BibitemShut {NoStop}%
\bibitem [{\citenamefont {Schmid}\ \emph {et~al.}(2025)\citenamefont {Schmid},
  \citenamefont {Peng}, \citenamefont {Refael},\ and\ \citenamefont {von
  Oppen}}]{schmid2025self}%
  \BibitemOpen
  \bibfield  {author} {\bibinfo {author} {\bibfnamefont {H.}~\bibnamefont
  {Schmid}}, \bibinfo {author} {\bibfnamefont {Y.}~\bibnamefont {Peng}},
  \bibinfo {author} {\bibfnamefont {G.}~\bibnamefont {Refael}},\ and\ \bibinfo
  {author} {\bibfnamefont {F.}~\bibnamefont {von Oppen}},\ }\bibfield  {title}
  {\bibinfo {title} {Self-similar phase diagram of the fibonacci-driven quantum
  ising model},\ }\href@noop {} {\bibfield  {journal} {\bibinfo  {journal}
  {Physical Review Letters}\ }\textbf {\bibinfo {volume} {134}},\ \bibinfo
  {pages} {240404} (\bibinfo {year} {2025})}\BibitemShut {NoStop}%
\bibitem [{\citenamefont {Dutta}\ \emph {et~al.}(2025)\citenamefont {Dutta},
  \citenamefont {Choudhury},\ and\ \citenamefont
  {Shukla}}]{dutta2025prethermalization}%
  \BibitemOpen
  \bibfield  {author} {\bibinfo {author} {\bibfnamefont {P.}~\bibnamefont
  {Dutta}}, \bibinfo {author} {\bibfnamefont {S.}~\bibnamefont {Choudhury}},\
  and\ \bibinfo {author} {\bibfnamefont {V.}~\bibnamefont {Shukla}},\
  }\bibfield  {title} {\bibinfo {title} {Prethermalization in the pxp model
  under continuous quasiperiodic driving},\ }\href@noop {} {\bibfield
  {journal} {\bibinfo  {journal} {Physical Review B}\ }\textbf {\bibinfo
  {volume} {111}},\ \bibinfo {pages} {064303} (\bibinfo {year}
  {2025})}\BibitemShut {NoStop}%
\bibitem [{\citenamefont {Kaushik}\ \emph {et~al.}(2026)\citenamefont
  {Kaushik}, \citenamefont {Choudhury},\ and\ \citenamefont {Nag}}]{prg5-3srn}%
  \BibitemOpen
  \bibfield  {author} {\bibinfo {author} {\bibfnamefont {V.}~\bibnamefont
  {Kaushik}}, \bibinfo {author} {\bibfnamefont {S.}~\bibnamefont {Choudhury}},\
  and\ \bibinfo {author} {\bibfnamefont {T.}~\bibnamefont {Nag}},\ }\bibfield
  {title} {\bibinfo {title} {Topological energy pumping in a quasiperiodically
  driven four-level system},\ }\href {https://doi.org/10.1103/prg5-3srn}
  {\bibfield  {journal} {\bibinfo  {journal} {Phys. Rev. B}\ }\textbf {\bibinfo
  {volume} {113}},\ \bibinfo {pages} {184312} (\bibinfo {year}
  {2026})}\BibitemShut {NoStop}%
\bibitem [{\citenamefont {Nag}\ and\ \citenamefont
  {Rajak}(2021)}]{PhysRevB.104.134307}%
  \BibitemOpen
  \bibfield  {author} {\bibinfo {author} {\bibfnamefont {T.}~\bibnamefont
  {Nag}}\ and\ \bibinfo {author} {\bibfnamefont {A.}~\bibnamefont {Rajak}},\
  }\bibfield  {title} {\bibinfo {title} {Periodic and aperiodic dynamics of
  flat bands in diamond-octagon lattice},\ }\href
  {https://doi.org/10.1103/PhysRevB.104.134307} {\bibfield  {journal} {\bibinfo
   {journal} {Phys. Rev. B}\ }\textbf {\bibinfo {volume} {104}},\ \bibinfo
  {pages} {134307} (\bibinfo {year} {2021})}\BibitemShut {NoStop}%
\bibitem [{\citenamefont {Kundu}\ \emph {et~al.}(2025)\citenamefont {Kundu},
  \citenamefont {Nag},\ and\ \citenamefont {Rajak}}]{lhnv-1l4l}%
  \BibitemOpen
  \bibfield  {author} {\bibinfo {author} {\bibfnamefont {A.}~\bibnamefont
  {Kundu}}, \bibinfo {author} {\bibfnamefont {T.}~\bibnamefont {Nag}},\ and\
  \bibinfo {author} {\bibfnamefont {A.}~\bibnamefont {Rajak}},\ }\bibfield
  {title} {\bibinfo {title} {Statistical prethermalization in a randomly kicked
  many-body classical rotor system},\ }\href
  {https://doi.org/10.1103/lhnv-1l4l} {\bibfield  {journal} {\bibinfo
  {journal} {Phys. Rev. B}\ }\textbf {\bibinfo {volume} {112}},\ \bibinfo
  {pages} {064310} (\bibinfo {year} {2025})}\BibitemShut {NoStop}%
\bibitem [{\citenamefont {Martin}\ \emph {et~al.}(2017)\citenamefont {Martin},
  \citenamefont {Refael},\ and\ \citenamefont
  {Halperin}}]{martin2017topological}%
  \BibitemOpen
  \bibfield  {author} {\bibinfo {author} {\bibfnamefont {I.}~\bibnamefont
  {Martin}}, \bibinfo {author} {\bibfnamefont {G.}~\bibnamefont {Refael}},\
  and\ \bibinfo {author} {\bibfnamefont {B.}~\bibnamefont {Halperin}},\
  }\bibfield  {title} {\bibinfo {title} {Topological frequency conversion in
  strongly driven quantum systems},\ }\href@noop {} {\bibfield  {journal}
  {\bibinfo  {journal} {Physical Review X}\ }\textbf {\bibinfo {volume} {7}},\
  \bibinfo {pages} {041008} (\bibinfo {year} {2017})}\BibitemShut {NoStop}%
\bibitem [{\citenamefont {Peng}\ and\ \citenamefont
  {Refael}(2018)}]{peng2018topological}%
  \BibitemOpen
  \bibfield  {author} {\bibinfo {author} {\bibfnamefont {Y.}~\bibnamefont
  {Peng}}\ and\ \bibinfo {author} {\bibfnamefont {G.}~\bibnamefont {Refael}},\
  }\bibfield  {title} {\bibinfo {title} {Topological energy conversion through
  the bulk or the boundary of driven systems},\ }\href@noop {} {\bibfield
  {journal} {\bibinfo  {journal} {Physical Review B}\ }\textbf {\bibinfo
  {volume} {97}},\ \bibinfo {pages} {134303} (\bibinfo {year}
  {2018})}\BibitemShut {NoStop}%
\bibitem [{\citenamefont {Qi}\ \emph {et~al.}(2021)\citenamefont {Qi},
  \citenamefont {Refael},\ and\ \citenamefont {Peng}}]{qi2021universal}%
  \BibitemOpen
  \bibfield  {author} {\bibinfo {author} {\bibfnamefont {Z.}~\bibnamefont
  {Qi}}, \bibinfo {author} {\bibfnamefont {G.}~\bibnamefont {Refael}},\ and\
  \bibinfo {author} {\bibfnamefont {Y.}~\bibnamefont {Peng}},\ }\bibfield
  {title} {\bibinfo {title} {Universal nonadiabatic energy pumping in a
  quasiperiodically driven extended system},\ }\href@noop {} {\bibfield
  {journal} {\bibinfo  {journal} {Physical Review B}\ }\textbf {\bibinfo
  {volume} {104}},\ \bibinfo {pages} {224301} (\bibinfo {year}
  {2021})}\BibitemShut {NoStop}%
\bibitem [{\citenamefont {Quan}\ \emph {et~al.}(2006)\citenamefont {Quan},
  \citenamefont {Song}, \citenamefont {Liu}, \citenamefont {Zanardi},\ and\
  \citenamefont {Sun}}]{PhysRevLett.96.140604}%
  \BibitemOpen
  \bibfield  {author} {\bibinfo {author} {\bibfnamefont {H.~T.}\ \bibnamefont
  {Quan}}, \bibinfo {author} {\bibfnamefont {Z.}~\bibnamefont {Song}}, \bibinfo
  {author} {\bibfnamefont {X.~F.}\ \bibnamefont {Liu}}, \bibinfo {author}
  {\bibfnamefont {P.}~\bibnamefont {Zanardi}},\ and\ \bibinfo {author}
  {\bibfnamefont {C.~P.}\ \bibnamefont {Sun}},\ }\bibfield  {title} {\bibinfo
  {title} {Decay of loschmidt echo enhanced by quantum criticality},\ }\href
  {https://doi.org/10.1103/PhysRevLett.96.140604} {\bibfield  {journal}
  {\bibinfo  {journal} {Phys. Rev. Lett.}\ }\textbf {\bibinfo {volume} {96}},\
  \bibinfo {pages} {140604} (\bibinfo {year} {2006})}\BibitemShut {NoStop}%
\bibitem [{\citenamefont {Vasista}\ \emph {et~al.}(2026)\citenamefont
  {Vasista}, \citenamefont {Agrawal},\ and\ \citenamefont {Nag}}]{t7m4-1qc4}%
  \BibitemOpen
  \bibfield  {author} {\bibinfo {author} {\bibfnamefont {A.~A.}\ \bibnamefont
  {Vasista}}, \bibinfo {author} {\bibfnamefont {A.}~\bibnamefont {Agrawal}},\
  and\ \bibinfo {author} {\bibfnamefont {T.}~\bibnamefont {Nag}},\ }\bibfield
  {title} {\bibinfo {title} {Generation of concurrence in a generalized central
  spin model with a three-spin interacting environment},\ }\href
  {https://doi.org/10.1103/t7m4-1qc4} {\bibfield  {journal} {\bibinfo
  {journal} {Phys. Rev. E}\ }\textbf {\bibinfo {volume} {113}},\ \bibinfo
  {pages} {044106} (\bibinfo {year} {2026})}\BibitemShut {NoStop}%
\bibitem [{\citenamefont {Nag}\ \emph {et~al.}(2012)\citenamefont {Nag},
  \citenamefont {Divakaran},\ and\ \citenamefont {Dutta}}]{PhysRevB.86.020401}%
  \BibitemOpen
  \bibfield  {author} {\bibinfo {author} {\bibfnamefont {T.}~\bibnamefont
  {Nag}}, \bibinfo {author} {\bibfnamefont {U.}~\bibnamefont {Divakaran}},\
  and\ \bibinfo {author} {\bibfnamefont {A.}~\bibnamefont {Dutta}},\ }\bibfield
   {title} {\bibinfo {title} {Scaling of the decoherence factor of a qubit
  coupled to a spin chain driven across quantum critical points},\ }\href
  {https://doi.org/10.1103/PhysRevB.86.020401} {\bibfield  {journal} {\bibinfo
  {journal} {Phys. Rev. B}\ }\textbf {\bibinfo {volume} {86}},\ \bibinfo
  {pages} {020401(R)} (\bibinfo {year} {2012})}\BibitemShut {NoStop}%
\bibitem [{\citenamefont {Suzuki}\ \emph {et~al.}(2016)\citenamefont {Suzuki},
  \citenamefont {Nag},\ and\ \citenamefont {Dutta}}]{PhysRevA.93.012112}%
  \BibitemOpen
  \bibfield  {author} {\bibinfo {author} {\bibfnamefont {S.}~\bibnamefont
  {Suzuki}}, \bibinfo {author} {\bibfnamefont {T.}~\bibnamefont {Nag}},\ and\
  \bibinfo {author} {\bibfnamefont {A.}~\bibnamefont {Dutta}},\ }\bibfield
  {title} {\bibinfo {title} {Dynamics of decoherence: Universal scaling of the
  decoherence factor},\ }\href {https://doi.org/10.1103/PhysRevA.93.012112}
  {\bibfield  {journal} {\bibinfo  {journal} {Phys. Rev. A}\ }\textbf {\bibinfo
  {volume} {93}},\ \bibinfo {pages} {012112} (\bibinfo {year}
  {2016})}\BibitemShut {NoStop}%
\bibitem [{\citenamefont {Nag}(2016)}]{PhysRevE.93.062119}%
  \BibitemOpen
  \bibfield  {author} {\bibinfo {author} {\bibfnamefont {T.}~\bibnamefont
  {Nag}},\ }\bibfield  {title} {\bibinfo {title} {Excess energy and decoherence
  factor of a qubit coupled to a one-dimensional periodically driven spin
  chain},\ }\href {https://doi.org/10.1103/PhysRevE.93.062119} {\bibfield
  {journal} {\bibinfo  {journal} {Phys. Rev. E}\ }\textbf {\bibinfo {volume}
  {93}},\ \bibinfo {pages} {062119} (\bibinfo {year} {2016})}\BibitemShut
  {NoStop}%
\bibitem [{\citenamefont {Cucchietti}\ \emph {et~al.}(2005)\citenamefont
  {Cucchietti}, \citenamefont {Paz},\ and\ \citenamefont
  {Zurek}}]{PhysRevA.72.052113}%
  \BibitemOpen
  \bibfield  {author} {\bibinfo {author} {\bibfnamefont {F.~M.}\ \bibnamefont
  {Cucchietti}}, \bibinfo {author} {\bibfnamefont {J.~P.}\ \bibnamefont
  {Paz}},\ and\ \bibinfo {author} {\bibfnamefont {W.~H.}\ \bibnamefont
  {Zurek}},\ }\bibfield  {title} {\bibinfo {title} {Decoherence from spin
  environments},\ }\href {https://doi.org/10.1103/PhysRevA.72.052113}
  {\bibfield  {journal} {\bibinfo  {journal} {Phys. Rev. A}\ }\textbf {\bibinfo
  {volume} {72}},\ \bibinfo {pages} {052113} (\bibinfo {year}
  {2005})}\BibitemShut {NoStop}%
\bibitem [{\citenamefont {Biswas}\ and\ \citenamefont
  {Choudhury}(2026)}]{biswas2025floquetcentralspinmodel}%
  \BibitemOpen
  \bibfield  {author} {\bibinfo {author} {\bibfnamefont {H.}~\bibnamefont
  {Biswas}}\ and\ \bibinfo {author} {\bibfnamefont {S.}~\bibnamefont
  {Choudhury}},\ }\bibfield  {title} {\bibinfo {title} {Floquet central spin
  model: A platform to realize time crystals, entanglement steering, and
  multiparameter metrology},\ }\href {https://doi.org/10.1103/vhx7-gbpt}
  {\bibfield  {journal} {\bibinfo  {journal} {Phys. Rev. B}\ }\textbf {\bibinfo
  {volume} {114}},\ \bibinfo {pages} {L140303} (\bibinfo {year}
  {2026})}\BibitemShut {NoStop}%
\bibitem [{\citenamefont {Urbaszek}\ \emph {et~al.}(2013)\citenamefont
  {Urbaszek}, \citenamefont {Marie}, \citenamefont {Amand}, \citenamefont
  {Krebs}, \citenamefont {Voisin}, \citenamefont {Maletinsky}, \citenamefont
  {H\"ogele},\ and\ \citenamefont {Imamoglu}}]{RevModPhys.85.79}%
  \BibitemOpen
  \bibfield  {author} {\bibinfo {author} {\bibfnamefont {B.}~\bibnamefont
  {Urbaszek}}, \bibinfo {author} {\bibfnamefont {X.}~\bibnamefont {Marie}},
  \bibinfo {author} {\bibfnamefont {T.}~\bibnamefont {Amand}}, \bibinfo
  {author} {\bibfnamefont {O.}~\bibnamefont {Krebs}}, \bibinfo {author}
  {\bibfnamefont {P.}~\bibnamefont {Voisin}}, \bibinfo {author} {\bibfnamefont
  {P.}~\bibnamefont {Maletinsky}}, \bibinfo {author} {\bibfnamefont
  {A.}~\bibnamefont {H\"ogele}},\ and\ \bibinfo {author} {\bibfnamefont
  {A.}~\bibnamefont {Imamoglu}},\ }\bibfield  {title} {\bibinfo {title}
  {Nuclear spin physics in quantum dots: An optical investigation},\ }\href
  {https://doi.org/10.1103/RevModPhys.85.79} {\bibfield  {journal} {\bibinfo
  {journal} {Rev. Mod. Phys.}\ }\textbf {\bibinfo {volume} {85}},\ \bibinfo
  {pages} {79} (\bibinfo {year} {2013})}\BibitemShut {NoStop}%
\bibitem [{\citenamefont {Ashida}\ \emph {et~al.}(2019)\citenamefont {Ashida},
  \citenamefont {Shi}, \citenamefont {Schmidt}, \citenamefont {Sadeghpour},
  \citenamefont {Cirac},\ and\ \citenamefont {Demler}}]{PhysRevA.100.043618}%
  \BibitemOpen
  \bibfield  {author} {\bibinfo {author} {\bibfnamefont {Y.}~\bibnamefont
  {Ashida}}, \bibinfo {author} {\bibfnamefont {T.}~\bibnamefont {Shi}},
  \bibinfo {author} {\bibfnamefont {R.}~\bibnamefont {Schmidt}}, \bibinfo
  {author} {\bibfnamefont {H.~R.}\ \bibnamefont {Sadeghpour}}, \bibinfo
  {author} {\bibfnamefont {J.~I.}\ \bibnamefont {Cirac}},\ and\ \bibinfo
  {author} {\bibfnamefont {E.}~\bibnamefont {Demler}},\ }\bibfield  {title}
  {\bibinfo {title} {Efficient variational approach to dynamics of a spatially
  extended bosonic kondo model},\ }\href
  {https://doi.org/10.1103/PhysRevA.100.043618} {\bibfield  {journal} {\bibinfo
   {journal} {Phys. Rev. A}\ }\textbf {\bibinfo {volume} {100}},\ \bibinfo
  {pages} {043618} (\bibinfo {year} {2019})}\BibitemShut {NoStop}%
\bibitem [{\citenamefont {Dobrzyniecki}\ and\ \citenamefont
  {Tomza}(2023)}]{PhysRevA.108.052618}%
  \BibitemOpen
  \bibfield  {author} {\bibinfo {author} {\bibfnamefont {J.}~\bibnamefont
  {Dobrzyniecki}}\ and\ \bibinfo {author} {\bibfnamefont {M.}~\bibnamefont
  {Tomza}},\ }\bibfield  {title} {\bibinfo {title} {Quantum simulation of the
  central spin model with a rydberg atom and polar molecules in optical
  tweezers},\ }\href {https://doi.org/10.1103/PhysRevA.108.052618} {\bibfield
  {journal} {\bibinfo  {journal} {Phys. Rev. A}\ }\textbf {\bibinfo {volume}
  {108}},\ \bibinfo {pages} {052618} (\bibinfo {year} {2023})}\BibitemShut
  {NoStop}%
\bibitem [{\citenamefont {Schwartz}\ \emph {et~al.}(2018)\citenamefont
  {Schwartz}, \citenamefont {Scheuer}, \citenamefont {Tratzmiller},
  \citenamefont {Müller}, \citenamefont {Chen}, \citenamefont {Dhand},
  \citenamefont {Wang}, \citenamefont {Müller}, \citenamefont {Naydenov},
  \citenamefont {Jelezko},\ and\ \citenamefont {Plenio}}]{Schwartz_2018}%
  \BibitemOpen
  \bibfield  {author} {\bibinfo {author} {\bibfnamefont {I.}~\bibnamefont
  {Schwartz}}, \bibinfo {author} {\bibfnamefont {J.}~\bibnamefont {Scheuer}},
  \bibinfo {author} {\bibfnamefont {B.}~\bibnamefont {Tratzmiller}}, \bibinfo
  {author} {\bibfnamefont {S.}~\bibnamefont {Müller}}, \bibinfo {author}
  {\bibfnamefont {Q.}~\bibnamefont {Chen}}, \bibinfo {author} {\bibfnamefont
  {I.}~\bibnamefont {Dhand}}, \bibinfo {author} {\bibfnamefont {Z.-Y.}\
  \bibnamefont {Wang}}, \bibinfo {author} {\bibfnamefont {C.}~\bibnamefont
  {Müller}}, \bibinfo {author} {\bibfnamefont {B.}~\bibnamefont {Naydenov}},
  \bibinfo {author} {\bibfnamefont {F.}~\bibnamefont {Jelezko}},\ and\ \bibinfo
  {author} {\bibfnamefont {M.~B.}\ \bibnamefont {Plenio}},\ }\bibfield  {title}
  {\bibinfo {title} {Robust optical polarization of nuclear spin baths using
  hamiltonian engineering of nitrogen-vacancy center quantum dynamics},\
  }\bibfield  {journal} {\bibinfo  {journal} {Science Advances}\ }\textbf
  {\bibinfo {volume} {4}},\ \href {https://doi.org/10.1126/sciadv.aat8978}
  {10.1126/sciadv.aat8978} (\bibinfo {year} {2018})\BibitemShut {NoStop}%
\bibitem [{\citenamefont {Childress}\ \emph {et~al.}(2006)\citenamefont
  {Childress}, \citenamefont {Gurudev~Dutt}, \citenamefont {Taylor},
  \citenamefont {Zibrov}, \citenamefont {Jelezko}, \citenamefont {Wrachtrup},
  \citenamefont {Hemmer},\ and\ \citenamefont {Lukin}}]{childress2006coherent}%
  \BibitemOpen
  \bibfield  {author} {\bibinfo {author} {\bibfnamefont {L.}~\bibnamefont
  {Childress}}, \bibinfo {author} {\bibfnamefont {M.~V.}\ \bibnamefont
  {Gurudev~Dutt}}, \bibinfo {author} {\bibfnamefont {J.~M.}\ \bibnamefont
  {Taylor}}, \bibinfo {author} {\bibfnamefont {A.~S.}\ \bibnamefont {Zibrov}},
  \bibinfo {author} {\bibfnamefont {F.}~\bibnamefont {Jelezko}}, \bibinfo
  {author} {\bibfnamefont {J.}~\bibnamefont {Wrachtrup}}, \bibinfo {author}
  {\bibfnamefont {P.~R.}\ \bibnamefont {Hemmer}},\ and\ \bibinfo {author}
  {\bibfnamefont {M.~D.}\ \bibnamefont {Lukin}},\ }\bibfield  {title} {\bibinfo
  {title} {Coherent dynamics of coupled electron and nuclear spin qubits in
  diamond},\ }\href {https://doi.org/10.1126/science.1131871} {\bibfield
  {journal} {\bibinfo  {journal} {Science}\ }\textbf {\bibinfo {volume}
  {314}},\ \bibinfo {pages} {281} (\bibinfo {year} {2006})}\BibitemShut
  {NoStop}%
\bibitem [{\citenamefont {Niknam}\ \emph {et~al.}(2021)\citenamefont {Niknam},
  \citenamefont {Santos},\ and\ \citenamefont {Cory}}]{PhysRevLett.127.080401}%
  \BibitemOpen
  \bibfield  {author} {\bibinfo {author} {\bibfnamefont {M.}~\bibnamefont
  {Niknam}}, \bibinfo {author} {\bibfnamefont {L.~F.}\ \bibnamefont {Santos}},\
  and\ \bibinfo {author} {\bibfnamefont {D.~G.}\ \bibnamefont {Cory}},\
  }\bibfield  {title} {\bibinfo {title} {Experimental detection of the
  correlation r\'enyi entropy in the central spin model},\ }\href
  {https://doi.org/10.1103/PhysRevLett.127.080401} {\bibfield  {journal}
  {\bibinfo  {journal} {Phys. Rev. Lett.}\ }\textbf {\bibinfo {volume} {127}},\
  \bibinfo {pages} {080401} (\bibinfo {year} {2021})}\BibitemShut {NoStop}%
\bibitem [{\citenamefont {Kumar}\ \emph {et~al.}(2025)\citenamefont {Kumar},
  \citenamefont {Frantzeskakis},\ and\ \citenamefont
  {Barnes}}]{PhysRevB.111.014302}%
  \BibitemOpen
  \bibfield  {author} {\bibinfo {author} {\bibfnamefont {A.}~\bibnamefont
  {Kumar}}, \bibinfo {author} {\bibfnamefont {R.}~\bibnamefont
  {Frantzeskakis}},\ and\ \bibinfo {author} {\bibfnamefont {E.}~\bibnamefont
  {Barnes}},\ }\bibfield  {title} {\bibinfo {title} {Hilbert space
  fragmentation and subspace scar time-crystallinity in driven homogeneous
  central-spin models},\ }\href {https://doi.org/10.1103/PhysRevB.111.014302}
  {\bibfield  {journal} {\bibinfo  {journal} {Phys. Rev. B}\ }\textbf {\bibinfo
  {volume} {111}},\ \bibinfo {pages} {014302} (\bibinfo {year}
  {2025})}\BibitemShut {NoStop}%
\bibitem [{\citenamefont {Piga}\ \emph {et~al.}(2019)\citenamefont {Piga},
  \citenamefont {Lewenstein},\ and\ \citenamefont
  {Quach}}]{PhysRevE.99.032213}%
  \BibitemOpen
  \bibfield  {author} {\bibinfo {author} {\bibfnamefont {A.}~\bibnamefont
  {Piga}}, \bibinfo {author} {\bibfnamefont {M.}~\bibnamefont {Lewenstein}},\
  and\ \bibinfo {author} {\bibfnamefont {J.~Q.}\ \bibnamefont {Quach}},\
  }\bibfield  {title} {\bibinfo {title} {Quantum chaos and entanglement in
  ergodic and nonergodic systems},\ }\href
  {https://doi.org/10.1103/PhysRevE.99.032213} {\bibfield  {journal} {\bibinfo
  {journal} {Phys. Rev. E}\ }\textbf {\bibinfo {volume} {99}},\ \bibinfo
  {pages} {032213} (\bibinfo {year} {2019})}\BibitemShut {NoStop}%
\bibitem [{\citenamefont {He}\ \emph {et~al.}(2025)\citenamefont {He},
  \citenamefont {Ye}, \citenamefont {Gong}, \citenamefont {Yao}, \citenamefont
  {Liu}, \citenamefont {Murch}, \citenamefont {Yao},\ and\ \citenamefont
  {Zu}}]{PhysRevX.15.011055}%
  \BibitemOpen
  \bibfield  {author} {\bibinfo {author} {\bibfnamefont {G.}~\bibnamefont
  {He}}, \bibinfo {author} {\bibfnamefont {B.}~\bibnamefont {Ye}}, \bibinfo
  {author} {\bibfnamefont {R.}~\bibnamefont {Gong}}, \bibinfo {author}
  {\bibfnamefont {C.}~\bibnamefont {Yao}}, \bibinfo {author} {\bibfnamefont
  {Z.}~\bibnamefont {Liu}}, \bibinfo {author} {\bibfnamefont {K.~W.}\
  \bibnamefont {Murch}}, \bibinfo {author} {\bibfnamefont {N.~Y.}\ \bibnamefont
  {Yao}},\ and\ \bibinfo {author} {\bibfnamefont {C.}~\bibnamefont {Zu}},\
  }\bibfield  {title} {\bibinfo {title} {Experimental realization of discrete
  time quasicrystals},\ }\href {https://doi.org/10.1103/PhysRevX.15.011055}
  {\bibfield  {journal} {\bibinfo  {journal} {Phys. Rev. X}\ }\textbf {\bibinfo
  {volume} {15}},\ \bibinfo {pages} {011055} (\bibinfo {year}
  {2025})}\BibitemShut {NoStop}%
\bibitem [{\citenamefont {Mori}(2022)}]{Mori2022heating}%
  \BibitemOpen
  \bibfield  {author} {\bibinfo {author} {\bibfnamefont {T.}~\bibnamefont
  {Mori}},\ }\bibfield  {title} {\bibinfo {title} {Heating rates under fast
  periodic driving beyond linear response},\ }\href
  {https://doi.org/10.1103/PhysRevLett.128.050604} {\bibfield  {journal}
  {\bibinfo  {journal} {Phys. Rev. Lett.}\ }\textbf {\bibinfo {volume} {128}},\
  \bibinfo {pages} {050604} (\bibinfo {year} {2022})}\BibitemShut {NoStop}%
\bibitem [{\citenamefont {Zhou}\ \emph {et~al.}(2026)\citenamefont {Zhou},
  \citenamefont {Mo}, \citenamefont {Zhao},\ and\ \citenamefont
  {Lapierre}}]{zhou2026emergentselfsimilarquantumrevivals}%
  \BibitemOpen
  \bibfield  {author} {\bibinfo {author} {\bibfnamefont {X.-C.}\ \bibnamefont
  {Zhou}}, \bibinfo {author} {\bibfnamefont {L.-H.}\ \bibnamefont {Mo}},
  \bibinfo {author} {\bibfnamefont {H.}~\bibnamefont {Zhao}},\ and\ \bibinfo
  {author} {\bibfnamefont {B.}~\bibnamefont {Lapierre}},\ }\href
  {https://arxiv.org/abs/2606.05288} {\bibinfo {title} {Emergent self-similar
  quantum revivals in spiral drives}} (\bibinfo {year} {2026}),\ \Eprint
  {https://arxiv.org/abs/2606.05288} {arXiv:2606.05288 [quant-ph]} \BibitemShut
  {NoStop}%
\bibitem [{sup()}]{suppmat}%
  \BibitemOpen
  \href@noop {} {}\bibinfo {note} {See Supplemental Material for a detailed
  description of the time evolution without entanglement generation and the
  Clifford dynamics.}\BibitemShut {Stop}%
\bibitem [{\citenamefont {Lajci}\ \emph {et~al.}(2026)\citenamefont {Lajci},
  \citenamefont {O'Dell},\ and\ \citenamefont
  {Mumford}}]{Lajcitopologically2026}%
  \BibitemOpen
  \bibfield  {author} {\bibinfo {author} {\bibfnamefont {B.}~\bibnamefont
  {Lajci}}, \bibinfo {author} {\bibfnamefont {D.~H.~J.}\ \bibnamefont
  {O'Dell}},\ and\ \bibinfo {author} {\bibfnamefont {J.}~\bibnamefont
  {Mumford}},\ }\bibfield  {title} {\bibinfo {title} {Topologically protected
  bell-cat states in a simple spin model},\ }\href
  {https://doi.org/10.1103/y4g4-bksr} {\bibfield  {journal} {\bibinfo
  {journal} {Phys. Rev. A}\ }\textbf {\bibinfo {volume} {113}},\ \bibinfo
  {pages} {013325} (\bibinfo {year} {2026})}\BibitemShut {NoStop}%
\bibitem [{\citenamefont {Dooley}\ \emph {et~al.}(2013)\citenamefont {Dooley},
  \citenamefont {McCrossan}, \citenamefont {Harland}, \citenamefont {Everitt},\
  and\ \citenamefont {Spiller}}]{PhysRevA.87.052323}%
  \BibitemOpen
  \bibfield  {author} {\bibinfo {author} {\bibfnamefont {S.}~\bibnamefont
  {Dooley}}, \bibinfo {author} {\bibfnamefont {F.}~\bibnamefont {McCrossan}},
  \bibinfo {author} {\bibfnamefont {D.}~\bibnamefont {Harland}}, \bibinfo
  {author} {\bibfnamefont {M.~J.}\ \bibnamefont {Everitt}},\ and\ \bibinfo
  {author} {\bibfnamefont {T.~P.}\ \bibnamefont {Spiller}},\ }\bibfield
  {title} {\bibinfo {title} {Collapse and revival and cat states with an
  $n$-spin system},\ }\href {https://doi.org/10.1103/PhysRevA.87.052323}
  {\bibfield  {journal} {\bibinfo  {journal} {Phys. Rev. A}\ }\textbf {\bibinfo
  {volume} {87}},\ \bibinfo {pages} {052323} (\bibinfo {year}
  {2013})}\BibitemShut {NoStop}%
\bibitem [{\citenamefont {Giovannetti}\ \emph {et~al.}(2006)\citenamefont
  {Giovannetti}, \citenamefont {Lloyd},\ and\ \citenamefont
  {Maccone}}]{giovannetti2006quantum}%
  \BibitemOpen
  \bibfield  {author} {\bibinfo {author} {\bibfnamefont {V.}~\bibnamefont
  {Giovannetti}}, \bibinfo {author} {\bibfnamefont {S.}~\bibnamefont {Lloyd}},\
  and\ \bibinfo {author} {\bibfnamefont {L.}~\bibnamefont {Maccone}},\
  }\bibfield  {title} {\bibinfo {title} {Quantum metrology},\ }\href
  {https://doi.org/10.1103/PhysRevLett.96.010401} {\bibfield  {journal}
  {\bibinfo  {journal} {Phys. Rev. Lett.}\ }\textbf {\bibinfo {volume} {96}},\
  \bibinfo {pages} {010401} (\bibinfo {year} {2006})}\BibitemShut {NoStop}%
\bibitem [{\citenamefont {Giovannetti}\ \emph {et~al.}(2011)\citenamefont
  {Giovannetti}, \citenamefont {Lloyd},\ and\ \citenamefont
  {Maccone}}]{giovannetti2011advances}%
  \BibitemOpen
  \bibfield  {author} {\bibinfo {author} {\bibfnamefont {V.}~\bibnamefont
  {Giovannetti}}, \bibinfo {author} {\bibfnamefont {S.}~\bibnamefont {Lloyd}},\
  and\ \bibinfo {author} {\bibfnamefont {L.}~\bibnamefont {Maccone}},\
  }\bibfield  {title} {\bibinfo {title} {Advances in quantum metrology},\
  }\href {https://doi.org/10.1038/nphoton.2011.35} {\bibfield  {journal}
  {\bibinfo  {journal} {Nature photonics}\ }\textbf {\bibinfo {volume} {5}},\
  \bibinfo {pages} {222} (\bibinfo {year} {2011})}\BibitemShut {NoStop}%
\bibitem [{\citenamefont {Asthana}\ \emph {et~al.}(2026)\citenamefont
  {Asthana}, \citenamefont {Alexander}, \citenamefont {Srivastava},
  \citenamefont {Mahesh},\ and\ \citenamefont
  {Vinjanampathy}}]{asthana2026efficientmanybodyshadowmetrology}%
  \BibitemOpen
  \bibfield  {author} {\bibinfo {author} {\bibfnamefont {S.}~\bibnamefont
  {Asthana}}, \bibinfo {author} {\bibfnamefont {C.}~\bibnamefont {Alexander}},
  \bibinfo {author} {\bibfnamefont {A.~K.}\ \bibnamefont {Srivastava}},
  \bibinfo {author} {\bibfnamefont {T.~S.}\ \bibnamefont {Mahesh}},\ and\
  \bibinfo {author} {\bibfnamefont {S.}~\bibnamefont {Vinjanampathy}},\ }\href
  {https://arxiv.org/abs/2603.24035} {\bibinfo {title} {Efficient many-body
  shadow metrology via clifford lensing}} (\bibinfo {year} {2026}),\ \Eprint
  {https://arxiv.org/abs/2603.24035} {arXiv:2603.24035 [quant-ph]} \BibitemShut
  {NoStop}%
\bibitem [{\citenamefont {Biswas}\ and\ \citenamefont
  {Choudhury}(2025)}]{biswas2025discretetimecrystalsspins}%
  \BibitemOpen
  \bibfield  {author} {\bibinfo {author} {\bibfnamefont {H.}~\bibnamefont
  {Biswas}}\ and\ \bibinfo {author} {\bibfnamefont {S.}~\bibnamefont
  {Choudhury}},\ }\href {https://arxiv.org/abs/2505.13207} {\bibinfo {title}
  {Discrete time crystals in the spin-s central spin model}} (\bibinfo {year}
  {2025}),\ \Eprint {https://arxiv.org/abs/2505.13207} {arXiv:2505.13207
  [quant-ph]} \BibitemShut {NoStop}%
\bibitem [{\citenamefont {Zhao}\ \emph {et~al.}(2021)\citenamefont {Zhao},
  \citenamefont {Mintert}, \citenamefont {Moessner},\ and\ \citenamefont
  {Knolle}}]{PhysRevLett.126.040601}%
  \BibitemOpen
  \bibfield  {author} {\bibinfo {author} {\bibfnamefont {H.}~\bibnamefont
  {Zhao}}, \bibinfo {author} {\bibfnamefont {F.}~\bibnamefont {Mintert}},
  \bibinfo {author} {\bibfnamefont {R.}~\bibnamefont {Moessner}},\ and\
  \bibinfo {author} {\bibfnamefont {J.}~\bibnamefont {Knolle}},\ }\bibfield
  {title} {\bibinfo {title} {Random multipolar driving: Tunably slow heating
  through spectral engineering},\ }\href
  {https://doi.org/10.1103/PhysRevLett.126.040601} {\bibfield  {journal}
  {\bibinfo  {journal} {Phys. Rev. Lett.}\ }\textbf {\bibinfo {volume} {126}},\
  \bibinfo {pages} {040601} (\bibinfo {year} {2021})}\BibitemShut {NoStop}%
\bibitem [{\citenamefont {Mori}\ \emph {et~al.}(2021)\citenamefont {Mori},
  \citenamefont {Zhao}, \citenamefont {Mintert}, \citenamefont {Knolle},\ and\
  \citenamefont {Moessner}}]{mori2021rigorous}%
  \BibitemOpen
  \bibfield  {author} {\bibinfo {author} {\bibfnamefont {T.}~\bibnamefont
  {Mori}}, \bibinfo {author} {\bibfnamefont {H.}~\bibnamefont {Zhao}}, \bibinfo
  {author} {\bibfnamefont {F.}~\bibnamefont {Mintert}}, \bibinfo {author}
  {\bibfnamefont {J.}~\bibnamefont {Knolle}},\ and\ \bibinfo {author}
  {\bibfnamefont {R.}~\bibnamefont {Moessner}},\ }\bibfield  {title} {\bibinfo
  {title} {Rigorous bounds on the heating rate in thue-morse quasiperiodically
  and randomly driven quantum many-body systems},\ }\href@noop {} {\bibfield
  {journal} {\bibinfo  {journal} {Physical review letters}\ }\textbf {\bibinfo
  {volume} {127}},\ \bibinfo {pages} {050602} (\bibinfo {year}
  {2021})}\BibitemShut {NoStop}%
\bibitem [{\citenamefont {Kumar}\ and\ \citenamefont
  {Choudhury}(2024)}]{PhysRevE.110.064150}%
  \BibitemOpen
  \bibfield  {author} {\bibinfo {author} {\bibfnamefont {S.}~\bibnamefont
  {Kumar}}\ and\ \bibinfo {author} {\bibfnamefont {S.}~\bibnamefont
  {Choudhury}},\ }\bibfield  {title} {\bibinfo {title} {Prethermalization in
  aperiodically driven classical spin systems},\ }\href
  {https://doi.org/10.1103/PhysRevE.110.064150} {\bibfield  {journal} {\bibinfo
   {journal} {Phys. Rev. E}\ }\textbf {\bibinfo {volume} {110}},\ \bibinfo
  {pages} {064150} (\bibinfo {year} {2024})}\BibitemShut {NoStop}%
\bibitem [{\citenamefont {Ng}\ \emph {et~al.}(2021)\citenamefont {Ng},
  \citenamefont {Wenderoth}, \citenamefont {Seelam}, \citenamefont {Rabani},
  \citenamefont {Meyer}, \citenamefont {Thoss},\ and\ \citenamefont
  {Kolodrubetz}}]{PhysRevB.103.134201}%
  \BibitemOpen
  \bibfield  {author} {\bibinfo {author} {\bibfnamefont {N.}~\bibnamefont
  {Ng}}, \bibinfo {author} {\bibfnamefont {S.}~\bibnamefont {Wenderoth}},
  \bibinfo {author} {\bibfnamefont {R.~R.}\ \bibnamefont {Seelam}}, \bibinfo
  {author} {\bibfnamefont {E.}~\bibnamefont {Rabani}}, \bibinfo {author}
  {\bibfnamefont {H.-D.}\ \bibnamefont {Meyer}}, \bibinfo {author}
  {\bibfnamefont {M.}~\bibnamefont {Thoss}},\ and\ \bibinfo {author}
  {\bibfnamefont {M.}~\bibnamefont {Kolodrubetz}},\ }\bibfield  {title}
  {\bibinfo {title} {Localization dynamics in a centrally coupled system},\
  }\href {https://doi.org/10.1103/PhysRevB.103.134201} {\bibfield  {journal}
  {\bibinfo  {journal} {Phys. Rev. B}\ }\textbf {\bibinfo {volume} {103}},\
  \bibinfo {pages} {134201} (\bibinfo {year} {2021})}\BibitemShut {NoStop}%
\end{thebibliography}%
\end{document}


\title{Supplemental Material for ``Quantum state revivals, time quasicrystals, and entanglement steering in a Fibonacci-driven central-spin model"}

\author{Anushka Agrawal}
\altaffiliation{These authors contributed equally to this work.}
\affiliation{Department of Physics, BITS Pilani, Hyderabad Campus, Telangana 500078, India}
\author{Adithya A. Vasista}
\altaffiliation{These authors contributed equally to this work.}
\affiliation{Department of Physics, BITS Pilani, Hyderabad Campus, Telangana 500078, India}
\author{Tanay Nag}
\affiliation{Department of Physics, BITS Pilani, Hyderabad Campus, Telangana 500078, India}
\author{Sayan Choudhury}
\email{sayanchoudhury@hri.res.in}
\affiliation{Harish-Chandra Research Institute, Chhatnag Road, Jhunsi, Prayagraj (Allahabad) 211019, India}
\affiliation{Homi Bhabha National Institute, Training School Complex, Anushakti Nagar, Mumbai 400094, India}

\date{\today}

\maketitle

In this Supplemental Material, we provide a detailed analysis of the dynamics of the system at $\lambda = 2\pi$, where no entanglement is generated in Sec.~\ref{sec:noentanglement} and the Clifford Points in Sec.~\ref{sec:cliffordfull}.

\section{Dynamics at $\lambda = 2 \pi$}
\label{sec:noentanglement}

In this section, we provide a detailed argument for the lack of entanglement generation when $\lambda = 2 \pi$, for any value of the
transverse field $\theta$ and any bath size $\Nsat$. We start by noting that $H_{1}=-\lambda S^{z}_{c}\sum_{i}S^{z}_{i}$ and $S^{\gamma}=\sigma^{\gamma}/2$, leading to the Ising Unitary
\begin{equation}
  \Uint=e^{-iH_{1}}
  =\exp\!\Big[\tfrac{i\lambda}{4}\,\sigma^{z}_{c}
   \textstyle\sum_{i}\sigma^{z}_{i}\Big]
  =\prod_{i=1}^{\Nsat}
   \exp\!\big[\tfrac{i\lambda}{4}\sigma^{z}_{c}\sigma^{z}_{i}\big],
  \label{eq:Uint_full}
\end{equation}
We note that each factor has the form $e^{i\chi P}$ with $P=\sigma^{z}_{c}\sigma^{z}_{i}$ and $P^{2}=\openone$, so that
$e^{i\chi P}=\cos\chi+i\sin\chi\,P$. At $\lambda=2\pi$ we have
$\chi=\pi/2$, the cosine vanishes, and each factor collapses to a
product of two single-spin Pauli operators,
\begin{equation}
  \exp\!\big[\tfrac{i\pi}{2}\sigma^{z}_{c}\sigma^{z}_{i}\big]
  = i\,\sigma^{z}_{c}\sigma^{z}_{i}.
  \label{eq:onefactor}
\end{equation}
Consequently, we obtain that
\begin{equation}
  \Uint\big|_{\lambda=2\pi}
  = i^{\Nsat}\,(\sigma^{z}_{c})^{\Nsat}\prod_{i=1}^{\Nsat}\sigma^{z}_{i}.
  \label{eq:factorise_SM}
\end{equation}
We note that $(\sigma^{z}_{c})^{\Nsat}$ equals $\openone$ for even
$\Nsat$ and $\sigma^{z}_{c}$ for odd $\Nsat$. \\
 
Furthermore, the transverse kick is described by the Unitary
\begin{equation}
  \Urot=\exp\!\Big[-\tfrac{i\theta}{2}\Big(\textstyle\sum_{i}\sigma^{x}_{i}
        +\sigma^{x}_{c}\Big)\Big]
      = e^{-i\theta\sigma^{x}_{c}/2}\otimes
        \bigotimes_{i=1}^{\Nsat} e^{-i\theta\sigma^{x}_{i}/2},
\end{equation}
which is a product of single-spin rotations for every $\theta$. At
$\lambda=2\pi$, therefore, both drive unitaries are products of
single-spin operators, and so is any sequence formed from them. A product
of local unitaries cannot generate entanglement, and the
reduced density matrix of the central spin stays pure and
$S_{\mathrm{vN}}=0$ throughout, for all $\theta$ and all $\Nsat$.
 
\begin{figure*}
    \centering
    \includegraphics[width=0.8\textwidth]{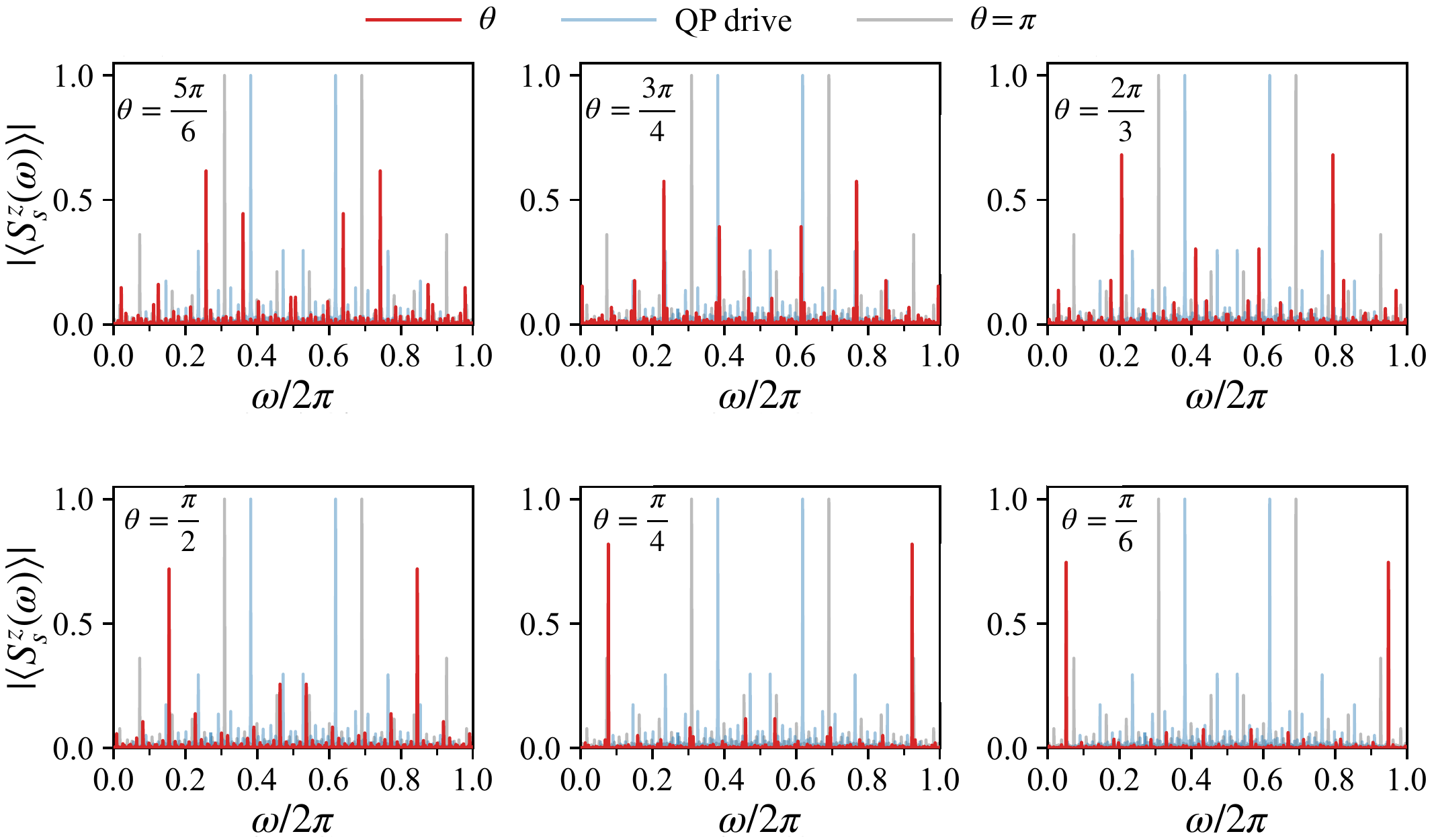}
    \caption{Fourier spectra of the satellite-spin magnetization,
$|\langle S_s^{z}(\omega)\rangle|$, when
$\lambda = 2\pi$ for decreasing transverse drive strengths
$\theta = 5\pi/6,\; 3\pi/4,\; 2\pi/3,\; \pi/2,\; \pi/4,\; \pi/6$.
The spectra for a perfect $\pi$-pulse ($\theta=\pi$) are overlaid
in red as a reference. The response shifts with changing $\theta$.}
    \label{fig:SuppFig1}
\end{figure*}

We note that the argument here does not depend on the order in which the Unitaries are applied. Thus, it applies not only to the Fibonacci drive studied here, but for any sequence built from the same two operations. Furthermore, although the entanglement is zero at stroboscopic times, the system exhibits quasiperiodic oscillations of the magnetization. The nature of these oscillations change with the strength of the transverse field as shown in Fig.~\ref{fig:SuppFig1}.

\section{Dynamics at Clifford Points}
\label{sec:cliffordfull}
\begin{figure*}[b]
    \centering
    \includegraphics[width=\textwidth]{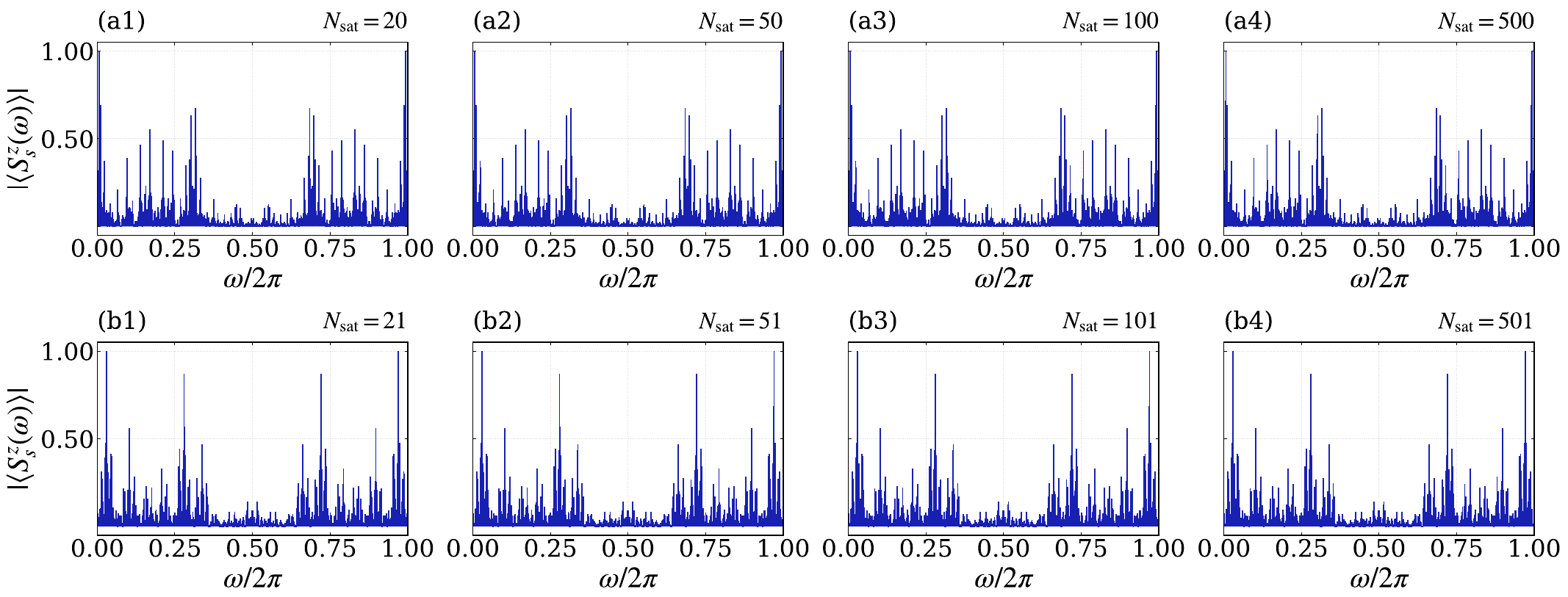}
    \caption{Fourier spectrum of the satellite spin magnetization $|\langle S^z_s(\omega)\rangle|$ for varying satellite spin bath sizes. Panels (a1)--(a4) correspond to even number of $N_{\mathrm{sat}} = 20, 50, 100,$ and $500$, and (b1)--(b4) correspond to odd number of $N_{sat} 21, 51, 101,$ and $501$  respectively. The system parameters are fixed at $\lambda = \pi$ and $\theta = \pi/2$. The location of the principal frequency peaks remains stable across all bath sizes, confirming that the subharmonic dynamical response is independent of finite-size effects.}
    \label{fig:fft_size_independence}
\end{figure*}

In this section, we analyze the CSM dynamics at the Clifford-points. We begin by noting that the nature of the oscillations depends on only on the parity of
$\Nsat$ (see Fig.~\ref{fig:fft_size_independence}), and that the revival periods can be obtained by composing two fixed permutations of a finite set. We begin our analysis by noting that at $\theta=\pi/2$, $\lambda=\pi$ the drive unitaries are
\begin{align}
  \Urot &= \exp\!\Big[-\tfrac{i\pi}{4}\Big(\textstyle\sum_{i}\sigma^{x}_{i}
          +\sigma^{x}_{c}\Big)\Big]
        = \bigotimes_{\text{all spins}} e^{-i\pi\sigma^{x}/4},
        \label{eq:Urot}\\[2pt]
  \Uint &= \exp\!\Big[\tfrac{i\pi}{4}\,\sigma^{z}_{c}
          \textstyle\sum_{i}\sigma^{z}_{i}\Big]
        = \prod_{i=1}^{\Nsat}
          \exp\!\big[\tfrac{i\pi}{4}\sigma^{z}_{c}\sigma^{z}_{i}\big],
        \label{eq:Uint}
\end{align}
a global $\sqrt{X}$ and a product of commuting $\sqrt{ZZ}$ gates, both
Clifford. The permutation symmetry of the model and the initial state ensures
that the generators never resolve individual satellites beyond one
distinguished site, so with
$\Ssat{\beta}\equiv\prod_{i}\sigma^{\beta}_{i}$ they retain the forms
\begin{equation}
  \mathcal{G}_{c}=(-1)^{a}\,\sigma^{\alpha}_{c}\,\Ssat{\beta},
  \qquad
  \mathcal{G}^{(j)}_{s}=(-1)^{b}\,\sigma^{\gamma}_{c}\,\sigma^{\delta}_{j}
  \!\!\prod_{i\neq j}\!\sigma^{\epsilon}_{i},
  \label{eq:alphabet}
\end{equation}
at every step, where $\sigma^{0}$ denotes the identity. We refer to the
Pauli indices as the \emph{labels} and to $a,b$ as the \emph{signs}.
\subsection{The two transition maps}
 
Conjugation by a single-spin $\sqrt{X}$ permutes the Pauli axes as
$\sigma^{x}\to\sigma^{x}$, $\sigma^{y}\to\sigma^{z}$,
$\sigma^{z}\to-\sigma^{y}$, with $\sigma^{0}$ invariant. For the Ising
gate we use $e^{i\pi P/4}Qe^{-i\pi P/4}=Q$ for $[P,Q]=0$ and $=iPQ$ for
$\{P,Q\}=0$. Applied to the $\Nsat$-fold string these give
Tables~\ref{tab:rot} and \ref{tab:int}.
 
\begin{table}
\caption{Action of $\Urot$ on the satellite string.}
\label{tab:rot}
\begin{ruledtabular}
\begin{tabular}{ccc}
 $\beta$ & $\Urot\,\Ssat{\beta}\,\Urot^{\dagger}$ & new $\beta$ \\ \hline
 $0$ & $\Ssat{0}$ & $0$ \\
 $x$ & $\Ssat{x}$ & $x$ \\
 $y$ & $\Ssat{z}$ & $z$ \\
 $z$ & $(-1)^{\Nsat}\,\Ssat{y}$ & $y$ \\
\end{tabular}
\end{ruledtabular}
\end{table}
 
\begin{table}
\caption{Action of $\Uint$ on the satellite string. The factor
$(\sigma^{z}_{c})^{\Nsat}$ is an operator, not a phase: it modifies the
central Pauli label.}
\label{tab:int}
\begin{ruledtabular}
\begin{tabular}{ccc}
 $\beta$ & $\Uint\,\Ssat{\beta}\,\Uint^{\dagger}$ & new $\beta$ \\ \hline
 $0$ & $\Ssat{0}$ & $0$ \\
 $x$ & $(-1)^{\Nsat}(\sigma^{z}_{c})^{\Nsat}\,\Ssat{y}$ & $y$ \\
 $y$ & $(\sigma^{z}_{c})^{\Nsat}\,\Ssat{x}$ & $x$ \\
 $z$ & $\Ssat{z}$ & $z$ \\
\end{tabular}
\end{ruledtabular}
\end{table}

The central factor transforms under $\Uint$ as
$\sigma^{z}_{c}\to\sigma^{z}_{c}$, $\sigma^{0}_{c}\to\sigma^{0}_{c}$,
and
\begin{equation}
  \sigma^{x}_{c}\longrightarrow
  (-i)^{\Nsat}\,\sigma^{x}_{c}(\sigma^{z}_{c})^{\Nsat}\,\Ssat{z},
  \label{eq:central}
\end{equation}
with the analogous expression for $\sigma^{y}_{c}$. This is the step at
which a satellite string is created from a bare central generator, and
it couples the central label back into the string label: the two are
not independent. Three crucial properties are evident from these considerations:

\begin{enumerate}
\item{The new labels are
determined entirely by the old labels; the signs never appear on the
right-hand side of a label transition. The labels therefore evolve as
a closed dynamical system, and the label period necessarily divides the
full period.}
 
\item{The only $\Nsat$-dependent objects generated are $(-1)^{\Nsat}$, $(-i)^{\Nsat}$
and $(\sigma^{z}_{c})^{\Nsat}$. The first two are scalars and affect
the signs only; $(\sigma^{z}_{c})^{\Nsat}$ equals $\openone$ for even
$\Nsat$ and $\sigma^{z}_{c}$ for odd $\Nsat$, and modifies the central
Pauli label $\alpha$. Although the direct $\beta$ transitions in
Tabs~\ref{tab:rot} and \ref{tab:int} are themselves
$\Nsat$-independent, the string label is not. The parity-dependent central label feeds back into $\beta$ through Eq.~\eqref{eq:central}, so both labels are in general parity dependent}
 
\item{The central generator carries a string over all $\Nsat$ satellites and accumulates $(\sigma^{z}_{c})^{\Nsat}$; the satellite generator carries a string over $\Nsat-1$ sites together with a distinguished site, and
accumulates $(\sigma^{z}_{c})^{\Nsat-1}$ from the string, with a
further factor of $\sigma^{z}_{c}$ when the label $\delta$ at the
distinguished site anticommutes with $\sigma^{z}$.}
\end{enumerate}
 
\subsection{Revival periods from the recursion}
 
Each drive unitary acts on the alphabet in Eq.~\eqref{eq:alphabet} as a fixed permutation: $\Lambda_{\mathrm{rot}}$ and $\Lambda_{\mathrm{int}}$ act
on the $16$ labels, or on the $64$ label-and-sign pairs when the signs
are retained. The Fibonacci construction lifts to the same recursion
on these permutations,
\begin{equation}
  \Lambda_{n}=\Lambda_{n-2}\circ\Lambda_{n-1},
  \qquad
  \Lambda_{0}=\Lambda_{\mathrm{rot}},\quad
  \Lambda_{1}=\Lambda_{\mathrm{int}},
  \label{eq:maprec}
\end{equation}
so the revival periods follow from iterating Eq.~\eqref{eq:maprec} on a
finite set, with no reference to the quantum state and no dependence on
the bath size beyond its parity. Since the recursion is invertible,
$\Lambda_{n-2}=\Lambda_{n}\circ\Lambda_{n-1}^{-1}$, the sequence is
purely periodic. Carrying out the iteration gives
Tab.~\ref{tab:periods}.
 
\begin{table}
\caption{Periods in the Fibonacci index $n$, obtained by iterating
Eq.~\eqref{eq:maprec}. The entanglement entropy is governed by the
central generator, the satellite magnetization by the satellite
generators.}
\label{tab:periods}
\begin{ruledtabular}
\begin{tabular}{lcc}
  & even $\Nsat$ & odd $\Nsat$ \\ \hline
 $\mathcal{G}_{c}$, labels only & $6$ & $6$ \\
 $\mathcal{G}^{(j)}_{s}$, labels only & $6$ & $6$ \\
 $\mathcal{G}^{(j)}_{s}$, labels and signs & $18$ & $12$ \\
\end{tabular}
\end{ruledtabular}
\end{table}
 
The label sector has period $6$ for both generator types and both
parities; retaining the signs multiplies this by $3$ for even $\Nsat$
and by $2$ for odd $\Nsat$.
 
\subsection{Consequences for the observables}
 
The satellite magnetization $\langle S^{z}_{s}\rangle$ is fixed by the
satellite generators and requires the sign of the corresponding group
element, since $\langle P\rangle=\pm1$ according to whether $+P$ or
$-P$ belongs to the stabilizer group. Its revival period in the
Fibonacci index is therefore $18$ for even $\Nsat$ and $12$ for odd
$\Nsat$. The entanglement entropy, on the other hand, is insensitive to signs and consequently, both parities exhibit period-6 oscillations in Fibonacci time. We note, the entropy sequences of the two parities coincide, not merely
in period. To see this, we note that at consecutive Fibonacci indices,
\begin{equation}
\begin{aligned}
  \text{even }\Nsat:&\quad
   (\alpha,\beta)=\ YI,\ YZ,\ ZI,\ YZ,\ YI,\ ZI,\\
  \text{odd }\Nsat:&\quad
   (\alpha,\beta)=\ YI,\ XZ,\ ZI,\ XZ,\ YI,\ ZI,
\end{aligned}
\label{eq:onecycle}
\end{equation}
so that the central labels differ while the string labels agree
throughout, leading to the same value of entanglement. By periodicity, the agreement in Eq.~\eqref{eq:onecycle} extends to all $n$. We note that this argument applies only at Fibonacci times, and parity-dependence can be seen for the stroboscopic entanglement oscillations, as seen in Fig.~4 of the main text. This observation is crucial, since it implies that the spin-cat states that arise for odd $\Nsat$, are generated at non-Fibonacci times only.